\documentclass[11pt,a4paper]{article}
\usepackage{amsmath,amssymb,amsfonts,amsthm,mathtools,bm}
\usepackage{booktabs,tabularx,enumitem,microtype}
\usepackage[round,authoryear]{natbib}
\usepackage[colorlinks=true,linkcolor=blue!55!black,citecolor=blue!55!black,urlcolor=blue!55!black]{hyperref}
\usepackage{xcolor}

\hypersetup{bookmarksdepth=1}
\allowdisplaybreaks[2]
\numberwithin{equation}{section}
\newtheorem{theorem}{Theorem}[section]
\newtheorem{proposition}[theorem]{Proposition}
\newtheorem{corollary}[theorem]{Corollary}

\theoremstyle{definition}\newtheorem{definition}[theorem]{Definition}
\theoremstyle{remark}\newtheorem{remark}[theorem]{Remark}
\newcommand{\E}{\mathbb E}\newcommand{\Cov}{\operatorname{Cov}}\newcommand{\Var}{\operatorname{Var}}
\newcommand{\sgn}{\operatorname{sgn}}\newcommand{\ind}{\mathbf 1}
\newcommand{\T}{\mathsf T}\newcommand{\bL}{\boldsymbol\Lambda}\newcommand{\bH}{\boldsymbol H}
\newcommand{\bG}{\boldsymbol\Gamma}\newcommand{\bS}{\boldsymbol\Sigma}
\newcommand{\ba}{\boldsymbol a}\newcommand{\bb}{\boldsymbol b}\newcommand{\bc}{\boldsymbol c}

\title{Exact Covariance Structure and 
Efficient Computation for L-Comoments}
\author{\begin{tabular}{c}
Palakorn Seenoi$^{a}$\qquad Jeong-Soo Park$^{b,*}$\\[7pt]
\small $^{a}$Department of Statistics, Faculty of Science, Khon Kaen University\\
\small Khon Kaen 40002, Thailand\\[5pt]
\small $^{b}$Department of Statistics, Chonnam National University\\
\small Gwangju 61186, South Korea\\[5pt]
\small $^{*}$Corresponding author: \href{mailto:jsparkstat@gmail.com}{jsparkstat@gmail.com}
\end{tabular}}
\date{\today}
\begin{document}

\maketitle
\begin{abstract}
L-moments are useful summary statistics and so have been widely employed in statistical analysis. L-comoments provide a multivariate generalization of L-moments. Joint finite-sample covariance structure, which represents the sampling uncertainty, of L-comoments is less well understood than that of univariate L-moments. We thus derive exact finite-sample covariance formulas for sample L-comoments using explicit conditional U-statistic kernels. We further construct an asymptotic projection
estimator and an unbiased estimator of the finite-sample covariance matrix and develop positive-definite corrections. An efficient rank-prefix algorithm evaluates the full unbiased covariance estimator in $O(n^2)$ time without enumerating disjoint subsets. Numerical experiments demonstrate substantial computational savings. The results provide a basis for first-order covariance approximations for L-comoment ratios and for subsequent multivariate L-moment estimation.
\end{abstract}

\section{Introduction}\label{sec:introduction}
L-moments summarize distributional location, scale, and shape through linear combinations of order statistics and require only first moments for their existence \citep{Hosking1990}. Methods based on L-moments have been widely used for distributional analysis and parameter estimation, notably through the method of L-moments estimation. Compared with conventional moments, L-moments are generally less sensitive to outlying observations and sampling variability. In some distributional settings, L-moment estimators are computationally efficient and can also be competitive with likelihood-based estimators, particularly for small samples \citep[for example,][]{Hosking1990,SankarasubramanianSrinivasan1999,HoskingWallis1997,MurshedSeoPark2014, ShinShinPark2025JKSS}. These properties make an understanding of their sampling distributions and finite-sample uncertainty especially valuable.

For univariate L-moments, \citet{ElamirSeheult2004} derived exact finite-sample variance--covariance formulas and distribution-free unbiased estimators, including a weighted order-statistic implementation. They also considered approximate standard errors for ratios of sample L-moments. Their findings provide a useful basis for studying the finite-sample behavior of L-moment statistics, which cannot in general be fully characterized by asymptotic approximations alone.

As a multivariate generalization, \citet{SerflingXiao2007} introduced directional L-comoments and L-comoment matrices, developed concomitant and U-statistic representations, and investigated asymptotic properties. They also noted that general U-statistic theory provides a route to exact variances and unbiased covariance estimation. However, explicit joint finite-sample covariance formulas for second and third L-comoments, together with an efficient unbiased covariance estimator, were not developed there. To the best of our knowledge, a comparable explicit treatment has not subsequently been available. Motivated by the univariate work of \citet{ElamirSeheult2004}, we address this gap. The general U-statistic overlap principle that we use is standard; our contribution is its explicit application to L-comoments.

L-comoments and related multivariate L-moment methods have been used in regional flood analysis \citep{ChebanaOuarda2007,AkinsojiEtAl2026}, drought regionalization \citep{ZhangXiaoSinghLi2012}, bivariate drought modeling \citep{AbdiEtAl2017}, and rainfall-frequency homogeneity studies \citep{ShinJeongJooHeo2018}. Related multivariate methods arise in homogeneity testing under spatial dependence \citep{Simkova2017}, joint marine extremes \citep{BaiRuanWang2023}, and process monitoring \citep{ArslanEtAl2025}. In these applications, the sample L-comoment matrix is used to summarize the dependence between the measured variables. In regional frequency analysis, for example, sample L-comoments and their ratios are compared across sites with short records, and their sampling variability is typically assessed by simulation from fitted parametric models \citep{ChebanaOuarda2007,Simkova2017}.
Our object of study is different: it is the \emph{sampling covariance
	matrix} of these estimated L-comoments, which quantifies their uncertainty. Throughout the paper, ``L-comoment'' refers to the population
quantity or its sample estimate, whereas ``sampling covariance matrix''
refers only to the covariance of the estimators.

Relative to \citet{SerflingXiao2007} and \citet{ElamirSeheult2004}, the main new contributions of this study are threefold.
\begin{description}[leftmargin=0pt,font=\normalfont\itshape,itemsep=0pt]
\item[First,] whereas \citet{SerflingXiao2007} obtained only the asymptotic covariance of the sample L-comoments, we give closed-form conditional kernels $M_{2,1}$, $M_{3,1}$, and $M_{3,2}$ for L-comoments, together with a canonical Hoeffding decomposition, so that every entry of the exact $4\times4$ and $6\times6$ covariance matrices is expressed through explicit expectations rather than through the general overlap identity alone. As a special case, this gives the exact joint covariance of the two directional Gini covariance estimators (Remark~\ref{rem:gini}).
\item[Second,] when $X$ and $Y$ are independent, we obtain a fully explicit covariance matrix.
\item[Third,] we show that the exactly unbiased covariance estimator can be
computed in $O(n^2)$ time, much faster than its direct evaluation, which
 requires $O(n^6)$ operations. The fast algorithm combines an overlap inclusion--exclusion identity with rank-prefix sums.
\end{description}

We also study a consistent positive-semidefinite (PSD) estimator of the leading asymptotic covariance; it also covers fourth-order L-comoments and is computed in $O(n\log n)$ time. In addition, we establish the first-order consistency of the unbiased estimator and give its positive-definite (PD) regularization. Numerical experiments confirm that the fast and direct algorithms agree
and compare their computing times. The experiments also show that
individual unbiased covariance estimates need not be PSD. These results are practically relevant because L-comoment-based regional
and bivariate frequency analyses often rely on short records, for example
a few decades of annual data \citep{HoskingWallis1997,ChebanaOuarda2007}.
For such sample sizes, asymptotic standard errors can be noticeably biased.
Section~\ref{sec:numerical} quantifies this bias and the accompanying
bias--variance trade-off between unbiased and asymptotic covariance
estimators.

The exact covariances of the L-comoments can also be used to approximate
the covariances of L-comoment ratios by a first-order Taylor
expansion, as in the delta method \citep{Oehlert1992,vanDerVaart1998};
these approximations are given in Appendix~\ref{subsec:ratios}.
Parameter estimation using L-comoments is not developed here. Related work uses L-moment methods for copula parameter estimation \citep{BrahimiChebanaNecir2015}, but copula modeling and parameter estimation are beyond the scope of the present study. Sample L-comoments can already be computed using functions such as \texttt{Lcomoment.Lk12} and \texttt{Lcomoment.matrix} in the \textsf{R} package \texttt{lmomco} \citep{Asquith2026}.

Section~\ref{sec:definitions} introduces L-comoments and their sample representations. Section~\ref{sec:general} derives the exact covariance structure. Section~\ref{sec:unbiased} develops covariance estimation, and Section~\ref{sec:fast} gives the efficient computation of the covariance estimators and their PSD and PD corrections. Section~\ref{sec:numerical} reports numerical experiments, Section~\ref{sec:discussion} discusses implications and extensions, and Section~\ref{sec:conclusion} concludes. The appendices collect background facts, derivations, computational details, and the covariance of L-comoment ratios.

\section{Directional L-comoments and sample constructions}\label{sec:definitions}

\subsection{Definitions}
Let $Z_b=(X_b,Y_b)$, $b=1,\ldots,n$, be independent and identically
distributed (iid) pairs from a continuous bivariate distribution. The two
coordinates of each pair may be dependent. We refer to $X$ and $Y$ as the
marginal variables, or simply the margins, and to $F_X$ and $F_Y$ as
their marginal distribution functions. We write $\mu_X=\E(X)$ and
$\mu_Y=\E(Y)$, and $Z=(X,Y)$ denotes a generic pair with the same
distribution as $Z_b$. Assume $\E(X^2+Y^2)<\infty$
whenever covariance is considered.

For $m=2,3,4$, the $m$th L-moment of $X$ \citep{Hosking1990} can be written as
\begin{equation}\label{eq:unilmom}
\lambda_m(X)=\E(X P_{m-1}\{F_X(X)\}),
\end{equation}
where $P_{m-1}$ is the shifted Legendre polynomial of degree $m-1$ on
$[0,1]$, given by
\begin{equation}\label{eq:polys}
P_1(u)=2u-1,\qquad P_2(u)=6u^2-6u+1,\qquad P_3(u)=20u^3-30u^2+12u-1.
\end{equation}
Here $\lambda_2(X)$ is the L-scale of $X$, $\lambda_3(X)$ measures the skewness of $X$, and $\lambda_4(X)$ measures the kurtosis of $X$.
The ratios $\lambda_3(X)/\lambda_2(X)$ and $\lambda_4(X)/\lambda_2(X)$ are the L-skewness and the L-kurtosis of $X$ \citep{Hosking1990}.

Replacing $F_X(X)$ by $F_Y(Y)$ in \eqref{eq:unilmom} defines the L-comoments
of \citet{SerflingXiao2007}, a bivariate extension of L-moments.

\begin{definition}[Directional L-comoments]\label{def:Lcomoment}
	For $m=2,3,4$, the $m$th directional L-comoment of $X$ with respect to
	$Y$ and that of $Y$ with respect to $X$ are
	\begin{equation}\label{eq:poplambda}
	\lambda_m^{[XY]}=\E(X P_{m-1}\{F_Y(Y)\}),\qquad
	\lambda_m^{[YX]}=\E(Y P_{m-1}\{F_X(X)\}).
	\end{equation}
	In the $[XY]$ direction, that is,
	\begin{align*}
	\lambda_2^{[XY]}&=\E[X\{2F_Y(Y)-1\}],\\
	\lambda_3^{[XY]}&=\E[X\{6F_Y(Y)^2-6F_Y(Y)+1\}],\\
	\lambda_4^{[XY]}&=\E[X\{20F_Y(Y)^3-30F_Y(Y)^2+12F_Y(Y)-1\}],
	\end{align*}
	and the $[YX]$ direction is obtained by exchanging $X$ and $Y$.
\end{definition}

The two directional L-comoments $\lambda_m^{[XY]}$ and $\lambda_m^{[YX]}$ are not equal in general, which is why we call them ``directional'' (asymmetric in the terminology of \citet{SerflingXiao2007}).
Since only directional L-comoments are considered in this paper, we
simply call them L-comoments from now on; the direction is indicated
by the superscript $[XY]$ or $[YX]$.

\subsection{Representation and interpretation of L-comoments}\label{subsec:represent}
L-comoments can also be written in terms of concomitants of order
statistics, in the same way as univariate L-moments are written in
terms of order statistics \citep{Hosking1990}. Consider a conceptual sample
$(X_1,Y_1),\ldots,(X_k,Y_k)$ of size $k$ from the distribution of $(X,Y)$.
If we order the sample by the $Y$ values and obtain the order statistics
$Y_{1:k}<\cdots<Y_{k:k}$, then the $X$ value paired with $Y_{r:k}$ is called
the \emph{concomitant} of $Y_{r:k}$ and is denoted by $X_{[r:k]}$
\citep{Yang1977,DavidNagaraja2003,JamalizadehBalakrishnan2012}.
Note that $X_{[r:k]}$ is not the $r$th order statistic $X_{r:k}$ of the
$X$ values. It is the $X$ value of the pair whose $Y$ value has rank $r$,
so $X_{[1:k]},\ldots,X_{[k:k]}$ are in general not in increasing order.
By Proposition~6 of \citet{SerflingXiao2007},
\begin{equation}\label{eq:popconcomitant}
\begin{aligned}
\lambda_2^{[XY]}&=\tfrac12\E\bigl(X_{[2:2]}-X_{[1:2]}\bigr),\\
\lambda_3^{[XY]}&=\tfrac13\E\bigl(X_{[3:3]}-2X_{[2:3]}+X_{[1:3]}\bigr),\\
\lambda_4^{[XY]}&=\tfrac14\E\bigl(X_{[4:4]}-3X_{[3:4]}+3X_{[2:4]}-X_{[1:4]}\bigr).
\end{aligned}
\end{equation}
In Appendix~\ref{app:kernelmean}, \eqref{eq:popconcomitant} is derived
from Definition~\ref{def:Lcomoment}. Thus, if two pairs are drawn and
the pair with the larger $Y$ value is identified, $\lambda_2^{[XY]}$ is
half the expected difference between the $X$ value of that pair and
the $X$ value of the other pair. Similarly, if three pairs are drawn
and sorted by their $Y$ values, $\lambda_3^{[XY]}$ compares the $X$
value of the middle pair with the $X$ values of the two outer pairs.

\citet{SerflingXiao2007} called $\lambda_2^{[XY]}$, $\lambda_3^{[XY]}$,
and $\lambda_4^{[XY]}$ the L-covariance, the L-coskewness, and the
L-cokurtosis. They called the standardized versions
$\tau_m^{[XY]}=\lambda_m^{[XY]}/\lambda_2(X)$ and
$\tau_m^{[YX]}=\lambda_m^{[YX]}/\lambda_2(Y)$ the L-comoment
coefficients, and $\tau_2^{[XY]}$ in particular the L-correlation.
The L-correlation lies in $[-1,1]$, and it equals $\pm1$ if and only if
$X$ is a strictly monotone function of $Y$
\citep[Proposition~4]{SerflingXiao2007}. Under linear regression of $X$
on $Y$ with affinely equivalent margins (that is, $Y$ has the same
distribution as $\alpha+\beta X$ for some $\alpha$ and some $\beta>0$), as in the bivariate normal
distribution, it equals the Pearson correlation
\citep[Proposition~3]{SerflingXiao2007}, but it requires only finite
first moments. Replacing $X$ by $F_X(X)$ in $\tau_2^{[XY]}$ gives
$12\Cov\{F_X(X),F_Y(Y)\}$, which is Spearman's rank correlation.
Each L-comoment carries the unit of its first variable, so it is divided
by the L-scale of that variable. At order two, these coefficients are
related to directional Gini correlations, and several symmetric Gini
constructions have also been studied
\citep{SangDangSang2016,VanderfordSangDang2020}. In
the rest of this paper we restrict attention to $m=2,3$, which are the
orders most frequently used in applications. The fourth L-comoment is
included above for completeness; Remark~\ref{rem:order4} and
Section~\ref{sec:discussion} discuss the extension of our results to
order four and beyond.

L-comoments are rank-based analogues of conventional
cross-moments. In particular, the second L-comoment is
\begin{equation*}
\lambda_2^{[XY]}=2\Cov\{X,F_Y(Y)\},
\end{equation*}
which is twice the Gini covariance of $X$ with respect to $Y$
\citep{SchechtmanYitzhaki1987,YitzhakiSchechtman2013,CarceaSerfling2015}. The third
L-comoment is
\begin{equation*}
\lambda_3^{[XY]}=6\Cov\{X,W\},\qquad W=\{F_Y(Y)-1/2\}^2 .
\end{equation*}
It is a rank-based analogue of the coskewness
$\E\{(X-\mu_X)(Y-\mu_Y)^2\}$ \citep{Rubinstein1973,HarveySiddique2000},
obtained by replacing $Y-\mu_Y$ with $F_Y(Y)-1/2$
\citep[Section 3.1.4]{SerflingXiao2007}.
Since $F_Y(Y)$ is uniform on $(0,1)$, $\E(W)=1/12$, and therefore
\begin{equation*}
\lambda_3^{[XY]}=6\{\E(XW)-\E(X)\E(W)\}=6\,\E\{X(W-1/12)\}.
\end{equation*}
The weight $W-1/12$ is positive when $F_Y(Y)<0.21$ or $F_Y(Y)>0.79$,
that is, when $Y$ lies in one of its two tails.
The weight is negative when $Y$ lies in the central part of its distribution.
Hence $\lambda_3^{[XY]}>0$ when $X$ is on average larger for values of $Y$ in its tails
than for values of $Y$ near its median, and $\lambda_3^{[XY]}<0$ in the opposite case.
For example, $\lambda_3^{[XY]}=0$ for the bivariate normal distribution,
because $\E(X\mid Y)$ is linear in $Y$ and $W$ is symmetric about the median of $Y$.
This is why $\lambda_3^{[XY]}$ plays the role of a coskewness: like
$\E\{(X-\mu_X)(Y-\mu_Y)^2\}$, it is positive when large values of $X$
occur with extreme values of $Y$ in either direction.
Unlike the univariate L-skewness, the L-coskewness $\lambda_3^{[XY]}$ is
not unit-free: it has the units of $X$ and is invariant to monotone
transformations of $Y$. However, the ratio
$\tau_3^{[XY]}=\lambda_3^{[XY]}/\lambda_2(X)$ is unit-free, because its
numerator and its denominator both have the units of $X$.

When $Y=X$, the L-comoments reduce to the univariate L-moments.
The Gini covariance has been studied extensively as a single measure
\citep{YitzhakiSchechtman2013}. In applications, however, L-comoments of
several orders are used jointly, for example in multivariate regional
frequency analysis \citep{ChebanaOuarda2007,AkinsojiEtAl2026} and copula
estimation \citep{BrahimiChebanaNecir2015}. Such joint use requires the
joint sampling covariance of the estimated L-comoments, which is the
subject of this paper.

\subsection{Estimation of  L-comoments}
To estimate $\lambda_m^{[XY]}$ from a sample, we follow the approach of
\citet{Hosking1990} for univariate L-moments. Each univariate L-moment is
the expectation of a simple function of a conceptual sample of size $m$.
For example, $\lambda_2(X)=\tfrac12\E(X_{2:2}-X_{1:2})$. For two
observations $x_i$ and $x_j$, the corresponding function is
$\tfrac12(x_{2:2}-x_{1:2})=\tfrac12|x_i-x_j|$. Averaging it over all
pairs of observations gives the unbiased estimator
\begin{equation*}
\widehat\lambda_2(X)={n\choose 2}^{-1}\sum_{i<j}\tfrac12|x_i-x_j| .
\end{equation*}
Higher-order sample L-moments are obtained in the same way from
subsamples of size $m$ \citep{Hosking1990}.

L-comoments are estimated by the same principle. The function that is
averaged over subsamples is called a kernel. Write $z_i=(x_i,y_i)$ for an
observed pair. A kernel of order $m$ takes $m$ distinct pairs
$z_1,\ldots,z_m$ and sorts them by their $y$ values. Let
$x_{[1]},\ldots,x_{[m]}$ be the corresponding $x$ values, that is, the
concomitants within these $m$ pairs. The kernel combines these
concomitants with the coefficients inside the expectations in
\eqref{eq:popconcomitant}. The order-two symmetric kernel is therefore
\begin{equation}\label{eq:Htwo}
H_2^{[XY]}(z_1,z_2)=\frac{x_{[2]}-x_{[1]}}2=\frac{(x_1-x_2)\sgn(y_1-y_2)}2,
\end{equation}
where $\sgn(t)$ equals $1$, $0$, or $-1$ according as $t>0$, $t=0$, or
$t<0$, and the order-three symmetric kernel is
\begin{equation}\label{eq:Hthree}
H_3^{[XY]}(z_1,z_2,z_3)=\frac{x_{[1]}-2x_{[2]}+x_{[3]}}3.
\end{equation}
The kernels $H_m^{[YX]}$ are obtained by exchanging the roles of the $x$
and $y$ coordinates. The kernel $H_2^{[XY]}$ lies between two familiar
kernels: replacing $\sgn(y_1-y_2)$ by $y_1-y_2$ in \eqref{eq:Htwo} gives
the kernel $(x_1-x_2)(y_1-y_2)/2$ of the sample covariance, and replacing
$x_1-x_2$ by $\sgn(x_1-x_2)$ gives the kernel
$\sgn(x_1-x_2)\sgn(y_1-y_2)$ of Kendall's tau. All these kernels are symmetric, that is, invariant
under permutations of their arguments, because the pairs are first
sorted by one coordinate. Symmetric kernels of this kind are standard in
U-statistic theory \citep{Hoeffding1948,Serfling1980,Lee1990}, which is the main tool of this paper; the facts used repeatedly are collected in Appendix~\ref{app:ustat}.

We now construct estimators of the L-comoments from the
sample. Since the two directions are estimated from the same pairs, we
treat them jointly in vector form. Write
\begin{equation*}
\boldsymbol\lambda_m=(\lambda_m^{[XY]},\lambda_m^{[YX]})^\T,\qquad
\bH_m=(H_m^{[XY]},H_m^{[YX]})^\T .
\end{equation*}
For an index set
$I=\{i_1,\ldots,i_m\}\subset\{1,\ldots,n\}$, write
$Z_I=(Z_{i_1},\ldots,Z_{i_m})$, so that
$\bH_m(Z_I)=\bH_m(Z_{i_1},\ldots,Z_{i_m})$; by symmetry, the order of
the indices does not matter.

For $m=2,3$ and $a\in\{[XY],[YX]\}$, the kernel $H_m^a$ evaluated at
any $m$ distinct pairs has mean $\lambda_m^a$; that is,
\begin{equation*}
\E\{\bH_m(Z_I)\}=\boldsymbol\lambda_m .
\end{equation*}
This follows from
\eqref{eq:popconcomitant}, and a proof is given in Appendix~\ref{app:kernelmean}.
We therefore estimate
$\boldsymbol\lambda_m$ by the complete U-statistic
\begin{equation}\label{eq:completeU}
\widehat{\boldsymbol\lambda}_m={n\choose m}^{-1}
\sum_{I\subset\{1,\ldots,n\},\,|I|=m}\bH_m(Z_I),\qquad m=2,3,
\end{equation}
which averages the kernel over all $\binom nm$ subsets of size $m$.
This estimator is unbiased for $\boldsymbol\lambda_m$. Each kernel is a
linear combination of $x$ values or of $y$ values, so it is square
integrable when $\E(X^2+Y^2)<\infty$. \citet{SerflingXiao2007} also gave
this U-statistic representation.

The sample means $\overline X$ and $\overline Y$ are U-statistics of
order one with kernels $H_1^{[XY]}(z)=x$ and $H_1^{[YX]}(z)=y$; they are
included in the covariance matrix of Section~\ref{sec:unbiased}.

The estimator \eqref{eq:completeU} also has a simple weighted-sum form.
Let $X_{[q:n]}$ be the concomitant of the $q$th smallest $Y$ value in the
sample; as in Section~\ref{sec:definitions}, $X_{[q:n]}$ is not the $q$th
order statistic of the $X$ values. Then the complete U-statistic \eqref{eq:completeU} can be written as
a weighted sum of the concomitants \citep[Section~3.3]{SerflingXiao2007}:
\begin{equation}\label{eq:rankstat}
\begin{gathered}
\widehat\lambda_m^{[XY]}=\sum_{q=1}^n w_{m,q}X_{[q:n]},\\
w_{2,q}=\frac{2q-n-1}{n(n-1)},\qquad
w_{3,q}=\frac{6q^2-6(n+1)q+(n+1)(n+2)}{n(n-1)(n-2)}.
\end{gathered}
\end{equation}
The weights follow by counting how often each pair occupies each
position in the kernels; Appendix~\ref{app:general} gives the details.
The weights $w_{2,q}$ require $n\ge2$, and the weights $w_{3,q}$ require
$n\ge3$. They satisfy $\sum_q w_{m,q}=0$. To make the estimating
equations \eqref{eq:completeU} and \eqref{eq:rankstat} more concrete,
Appendix~\ref{app:example} restates both in elementary terms, gives the
equivalent probability-weighted-moment form of \eqref{eq:rankstat}, and
evaluates both for a sample of size five.

We collect the four L-comoments and their estimators as
\begin{equation*}
\bL=(\lambda_2^{[XY]},\lambda_2^{[YX]},\lambda_3^{[XY]},\lambda_3^{[YX]})^\T,
\qquad
\widehat\bL=(\widehat\lambda_2^{[XY]},\widehat\lambda_2^{[YX]},\widehat\lambda_3^{[XY]},\widehat\lambda_3^{[YX]})^\T.
\end{equation*}
By \eqref{eq:completeU}, $\widehat\bL$ is unbiased for $\bL$.
\citet[Proposition~8]{SerflingXiao2007} showed that, under second-moment
assumptions, the sample L-comoments of one direction are asymptotically
multivariate normal. Their asymptotic covariance matrix is the
corresponding block of $B_1/n$ in Theorem~\ref{thm:Bdecomp} below. This approximation
can be poor for the sample sizes at which L-comoments are typically used.
In hydrology, for example, at-site records often contain only $20$ to
$50$ observations, and Section~\ref{subsec:reference} shows that the
asymptotic covariance then understates the finite-sample covariance.
Moreover, applications such as discordancy and homogeneity tests combine
L-comoments of several orders and both directions. They therefore require
the whole matrix $\Cov(\widehat\bL)$. In current practice, this matrix is
approximated by simulation from a fitted parametric model
\citep{ChebanaOuarda2007}. For these reasons, the main aim of this paper
is to derive the exact finite-sample form of $\Cov(\widehat\bL)$ for every
fixed $n$, without a parametric model, and to estimate this matrix
unbiasedly and efficiently from the data.
First-order covariance approximations for the ratios $\tau_m^{[XY]}$ and
$\tau_m^{[YX]}$ of Section~\ref{sec:definitions} are given in
Appendix~\ref{subsec:ratios}.

\section{Exact finite-sample covariance}\label{sec:general}

In this section we derive the exact finite-sample covariance matrix
$\Cov(\widehat\bL)$ for any fixed $n$, without assuming a parametric
form for the dependence between $X$ and $Y$. Theorem~\ref{thm:exact}
expresses this matrix through the conditional means of the kernels
defined below. Section~\ref{subsec:kernels} gives these conditional
means explicitly, and Section~\ref{subsec:indep} obtains the covariance
in closed form when $X$ and $Y$ are independent.

\paragraph{Overlap identity.}
Since sample L-comoments are U-statistics, their covariances follow from
a standard identity of U-statistic theory. The idea is as follows. Each
sample L-comoment is an average of kernel values over subsets of the
pairs. Consider one kernel value of $\widehat\lambda_m^a$, computed from
a subset $I$ of $m$ pairs, and one kernel value of
$\widehat\lambda_s^b$, computed from a subset $J$ of $s$ pairs. If $I$
and $J$ have no pair in common, the two kernel values are independent,
because the pairs are independent; hence their covariance is zero. If
$I$ and $J$ share $c\ge1$ pairs, the two kernel values are dependent only
through these $c$ shared pairs. Given the shared pairs, the two kernel
values are independent, so their covariance is the covariance of their
conditional means given the shared pairs.

We therefore define the conditional mean of a kernel given $c$ of its
pairs, which we call a conditional kernel. For $a\in\{[XY],[YX]\}$ and $1\le c\le m$, let
\begin{equation}\label{eq:Mm}
M_{m,c}^{a}(z_1,\ldots,z_c)
=\E(H_m^a(Z_1,\ldots,Z_m)\mid Z_1=z_1,\ldots,Z_c=z_c),
\end{equation}
which averages the kernel over the remaining $m-c$ pairs. In particular,
$M_{m,m}^a=H_m^a$. By the law of iterated expectations and
$\E\{H_m^a(Z_1,\ldots,Z_m)\}=\lambda_m^a$ (Section~\ref{sec:definitions}),
\begin{equation*}
\E\{M_{m,c}^a(Z_1,\ldots,Z_c)\}=\lambda_m^a .
\end{equation*}
The case $c=1$ plays a special role. The function $M_{m,1}^a(z)$ is the
mean of the kernel when one of its $m$ pairs is fixed at $z$, and it is
called the \emph{first projection} of the kernel $H_m^a$
\citep{Hoeffding1948,Serfling1980}. The name comes from the fact that
\begin{equation}\label{eq:projsum}
\frac mn\sum_{i=1}^n\{M_{m,1}^a(Z_i)-\lambda_m^a\}
\end{equation}
is the projection of $\widehat\lambda_m^a-\lambda_m^a$ onto sums of the
form $\sum_{i=1}^n g(Z_i)$, that is, the sum of functions of single
observations that is closest to $\widehat\lambda_m^a-\lambda_m^a$ in mean
square. This sum determines the leading term of the covariance of
$\widehat\lambda_m^a$ (Section~\ref{sec:unbiased}).
Let each of $a$ and $b$ denote one of the two directions, $[XY]$ or $[YX]$,
and let each of $m$ and $s$ denote one of the orders, $2$ or $3$.
For $1\le c\le\min(m,s)$, define
\begin{equation}\label{eq:gammas}
\Gamma_{ms,c}^{ab}
=\Cov\{M_{m,c}^{a}(Z_1,\ldots,Z_c),\,M_{s,c}^{b}(Z_1,\ldots,Z_c)\}
=\E(M_{m,c}^aM_{s,c}^b)-\lambda_m^a\lambda_s^b .
\end{equation}
By the argument above, $\Gamma_{ms,c}^{ab}$ is the covariance between a
kernel value of $\widehat\lambda_m^a$ and a kernel value of
$\widehat\lambda_s^b$ whose subsets share exactly $c$ pairs.
For fixed $m$, $s$, and $c$, we collect the four values of
$\Gamma_{ms,c}^{ab}$ in the $2\times2$ matrix
\begin{equation*}
\bG_{ms,c}=
\begin{pmatrix}
\Gamma_{ms,c}^{[XY][XY]} & \Gamma_{ms,c}^{[XY][YX]}\\[2pt]
\Gamma_{ms,c}^{[YX][XY]} & \Gamma_{ms,c}^{[YX][YX]}
\end{pmatrix}.
\end{equation*}
Its rows correspond to the direction of the order-$m$ kernel, and its
columns correspond to the direction of the order-$s$ kernel. The matrix
$\bG_{mm,c}$ is symmetric, but $\bG_{23,c}$ need not be symmetric.
These matrices are the building blocks of Theorem~\ref{thm:exact}: the
blocks of $\Cov(\widehat\bL)$ in \eqref{eq:exactblock} and
\eqref{eq:blocks} are linear combinations of them.

To combine these covariances, we count the pairs of subsets $(I,J)$ by
the number $c$ of shared pairs. Pairs with $c=0$ contribute nothing,
and each pair with $c\ge1$ contributes $\Gamma_{ms,c}^{ab}$. This gives the
\emph{overlap identity}
\citep[see][Chapter~5]{Hoeffding1948,Lee1990,Serfling1980}:
\begin{equation}\label{eq:overlapmaster}
\Cov(\widehat\lambda_m^a,\widehat\lambda_s^b)=\sum_{c=1}^{\min(m,s)}\omega_{ms,c}(n)\,\Gamma_{ms,c}^{ab},
\qquad
\omega_{ms,c}(n)=\frac{{m\choose c}{n-m\choose s-c}}{{n\choose s}},
\end{equation}
where $\omega_{ms,c}(n)$ is the proportion of subsets $J$ of size $s$
that share exactly $c$ pairs with a fixed subset $I$ of size $m$. The
identity holds for kernels of any fixed order.
\medskip

In this paper we apply the overlap identity \eqref{eq:overlapmaster} to
the second and third sample L-comoments, that is, to the U-statistics
with the order-two kernels $H_2^a$ in \eqref{eq:Htwo} and the
order-three kernels $H_3^a$ in \eqref{eq:Hthree}. The second L-comoment
$\lambda_2^{[XY]}$ is twice the Gini covariance of $X$ with respect to
$Y$. The third L-comoment $\lambda_3^{[XY]}$ measures whether $X$ is
larger when $Y$ lies in its tails than when $Y$ lies near its median
(Section~\ref{sec:definitions}). For these two orders, Theorem~\ref{thm:exact} below gives every entry
\begin{equation*}
\Cov(\widehat\lambda_m^a,\widehat\lambda_s^b),\qquad m,s\in\{2,3\},\quad a,b\in\{[XY],[YX]\},
\end{equation*}
of $\Cov(\widehat\bL)$ explicitly. These entries include covariances
between the two directions, such as
$\Cov(\widehat\lambda_2^{[XY]},\widehat\lambda_2^{[YX]})$, and covariances
between orders, such as
$\Cov(\widehat\lambda_2^{[XY]},\widehat\lambda_3^{[XY]})$.

We now state the main result of this section. It expresses the exact
covariance matrix of $\widehat\bL$ for every $n\ge3$ in terms of the
$2\times2$ matrices $\bG_{ms,c}$.
\begin{theorem}[Explicit joint covariance for orders two and three]\label{thm:exact}
For every $n\ge3$, the covariance of
$\widehat\bL$  
is exactly
\begin{equation}\label{eq:exactblock}
\Cov(\widehat\bL)=
\begin{pmatrix}\bS_{22,n}&\bS_{23,n}\\\bS_{23,n}^\T&\bS_{33,n}\end{pmatrix},
\end{equation}
where
\begin{align}\label{eq:blocks}
\bS_{22,n}&=\frac{4(n-2)}{n(n-1)}\bG_{22,1}+\frac2{n(n-1)}\bG_{22,2},\\
\bS_{23,n}&=\frac{6(n-3)}{n(n-1)}\bG_{23,1}+\frac6{n(n-1)}\bG_{23,2},\nonumber\\
\bS_{33,n}&=\frac{9(n-3)(n-4)}{n(n-1)(n-2)}\bG_{33,1}
+\frac{18(n-3)}{n(n-1)(n-2)}\bG_{33,2}
+\frac6{n(n-1)(n-2)}\bG_{33,3}.\nonumber
\end{align}
\end{theorem}
\begin{proof}
The identity \eqref{eq:overlapmaster} is standard (see \eqref{eq:appoverlap} in Appendix~\ref{app:ustat}), so only the coefficients $\omega_{ms,c}(n)$ need to be computed to complete the blocks \eqref{eq:blocks} of the covariance matrix \eqref{eq:exactblock}. For $(m,s)=(2,2)$, the overlap counts are $2(n-2)$ and $1$, with denominator $n(n-1)/2$, yielding $4(n-2)/\{n(n-1)\}$ and $2/\{n(n-1)\}$. For $(m,s)=(2,3)$, the counts are $2{n-2\choose2}=(n-2)(n-3)$ and $n-2$, with denominator ${n\choose3}=n(n-1)(n-2)/6$; these yield $6(n-3)/\{n(n-1)\}$ and $6/\{n(n-1)\}$. For $(m,s)=(3,3)$, the counts are $3{n-3\choose2}$, $3(n-3)$, and $1$, with denominator ${n\choose3}$, giving exactly the three coefficients in \eqref{eq:blocks}. Symmetry of the resulting covariance gives the lower-left block as the transpose of the upper-right block.
\end{proof}

Theorem~\ref{thm:exact} does not yet give the covariance in a form that
can be evaluated. It expresses $\Cov(\widehat\bL)$ through the matrices
$\bG_{ms,c}$, and each entry of $\bG_{ms,c}$ is a covariance of
conditional kernels. These conditional kernels are defined in
\eqref{eq:Mm} only as conditional expectations. To evaluate the
covariance, they must be written as explicit functions of their
arguments. For $c=m$ this is immediate, because $M_{m,m}^a=H_m^a$.
The remaining conditional kernels, $M_{2,1}$, $M_{3,2}$, and $M_{3,1}$,
are derived in Section~\ref{subsec:kernels}. With them, every entry of
$\bG_{ms,c}$ becomes the expectation of a known function of $(X,Y)$, as
illustrated by Corollary~\ref{cor:order2} and Theorem~\ref{thm:indep}.

\subsection{The conditional kernels needed in the covariance}\label{subsec:kernels}
We now give the conditional kernels $M_{2,1}$, $M_{3,2}$, and $M_{3,1}$
in closed form for the $[XY]$ direction. Write $\mu_X=\E(X)$, $u_y=F_Y(y)$, and
$Q_X(y)=\E(X\ind(Y\le y))$, where $\ind(A)$ denotes the indicator of the
event $A$. Then
\begin{equation}\label{eq:M21main}
M_{2,1}^{[XY]}(x,y)
=\frac12\{x(2u_y-1)+\mu_X-2Q_X(y)\}.
\end{equation}
For two fixed pairs with $y_\ell<y_h$, let $x_\ell$ and $x_h$ be the concomitants of $y_\ell$ and $y_h$, respectively, and put $u_\ell=F_Y(y_\ell),u_h=F_Y(y_h)$. Then
\begin{equation}\label{eq:M32main}
M_{3,2}^{[XY]}(z_1,z_2)
=\frac13\{(1-3u_\ell)x_\ell+(3u_h-2)x_h+\mu_X-3[Q_X(y_h)-Q_X(y_\ell)]\}.
\end{equation}
Finally let
\begin{align}\label{eq:RKmain}
R_X(y)&=\E(X P_2'(F_Y(Y))\ind(Y\ge y)),\\
K_X&=\E(X F_Y(Y)P_2'(F_Y(Y))).\nonumber
\end{align}
Then
\begin{equation}\label{eq:M31main}
M_{3,1}^{[XY]}(x,y)
=\frac13\{xP_2(u_y)+2\lambda_3^{[XY]}+R_X(y)-K_X\}.
\end{equation}
For the reverse direction, exchange $X$ and $Y$. All three expressions follow by conditioning on where the $y$ values of the remaining pairs fall relative to the given $y$ values; this determines the position of each pair in the ordering by $y$ and hence the coefficient of its $x$ value in the kernel. Detailed case-by-case derivations are given in Appendix~\ref{app:general}.

The simplest case is $m=s=c=2$, in which the conditional kernel is the kernel itself, $M_{2,2}^a=H_2^a$. The following corollary shows that $\bG_{22,2}$ then reduces to elementary moments of $(X,Y)$.

\begin{corollary}[Explicit order-two overlap matrix]\label{cor:order2}
Let $Z=(X,Y)$ and $Z'=(X',Y')$ be independent copies of the bivariate random vector. Then
\begin{equation}\label{eq:Gamma22two}
\bG_{22,2}=
\begin{pmatrix}
\frac12\Var(X)-(\lambda_2^{[XY]})^2&\frac14\E(|X-X'||Y-Y'|)-\lambda_2^{[XY]}\lambda_2^{[YX]}\\
\frac14\E(|X-X'||Y-Y'|)-\lambda_2^{[XY]}\lambda_2^{[YX]}&\frac12\Var(Y)-(\lambda_2^{[YX]})^2
\end{pmatrix}.
\end{equation}
\end{corollary}
\begin{proof}
By \eqref{eq:gammas} with $M_{2,2}^a=H_2^a$, the entries of $\bG_{22,2}$ are
$\Gamma_{22,2}^{ab}=\E\{H_2^a(Z,Z')H_2^b(Z,Z')\}-\lambda_2^a\lambda_2^b$.
Since $Y$ has a continuous distribution, $\Pr(Y=Y')=0$, so
$\{\sgn(Y-Y')\}^2=1$ with probability one. Hence
$\{H_2^{[XY]}(Z,Z')\}^2=(X-X')^2/4$ with probability one. Because
$\E\{(X-X')^2\}=2\Var(X)$, the expectation of $\{H_2^{[XY]}(Z,Z')\}^2$ is
$\Var(X)/2$. Subtracting $(\lambda_2^{[XY]})^2$ gives the upper-left entry of
\eqref{eq:Gamma22two}; exchanging $X$ and $Y$ gives the lower-right
entry. For the off-diagonal entries,
\begin{align*}
H_2^{[XY]}(Z,Z')H_2^{[YX]}(Z,Z')
&=\frac14(X-X')\sgn(X-X')\,(Y-Y')\sgn(Y-Y')\\
&=\frac14|X-X'||Y-Y'|.
\end{align*}
Taking expectations and subtracting $\lambda_2^{[XY]}\lambda_2^{[YX]}$ gives the
upper-right entry $\Gamma_{22,2}^{[XY][YX]}$. The lower-left entry
$\Gamma_{22,2}^{[YX][XY]}$ is the same, because the product is symmetric
in the two directions.
\end{proof}

\begin{remark}[Gini covariances]\label{rem:gini}
Since $\lambda_2^{[XY]}=2\Cov\{X,F_Y(Y)\}$, the estimator
$\widehat G^{[XY]}=\widehat\lambda_2^{[XY]}/2$ is the U-statistic
estimator of the Gini covariance $G^{[XY]}=\Cov\{X,F_Y(Y)\}$ of $X$ with
respect to $Y$ \citep{SchechtmanYitzhaki1987}; $\widehat G^{[YX]}$ and
$G^{[YX]}$ are defined by exchanging $X$ and $Y$.
Dividing $\bS_{22,n}$ in Theorem~\ref{thm:exact} by four and using
Corollary~\ref{cor:order2}, we obtain, for $a,b\in\{[XY],[YX]\}$ and the conditional kernels $M_{2,1}^a$ in \eqref{eq:M21main},
\begin{equation}\label{eq:ginicov}
\Cov(\widehat G^{a},\widehat G^{b})
=\frac{n-2}{n(n-1)}\Cov\{M_{2,1}^{a}(Z),M_{2,1}^{b}(Z)\}
+\frac{1}{2n(n-1)}\Gamma_{22,2}^{ab}.
\end{equation}
In particular,
\begin{align*}
\Var(\widehat G^{[XY]})
&=\frac{n-2}{n(n-1)}\Var\{M_{2,1}^{[XY]}(Z)\}
+\frac{1}{2n(n-1)}\Bigl\{\tfrac12\Var(X)-4(G^{[XY]})^2\Bigr\},\\
\Cov(\widehat G^{[XY]},\widehat G^{[YX]})
&=\frac{n-2}{n(n-1)}\Cov\{M_{2,1}^{[XY]}(Z),M_{2,1}^{[YX]}(Z)\}\\
&\quad+\frac{1}{2n(n-1)}\Bigl\{\tfrac14\E(|X-X'||Y-Y'|)-4G^{[XY]}G^{[YX]}\Bigr\}.
\end{align*}
When $X$ and $Y$ are independent, Theorem~\ref{thm:indep} below gives
$\Var(\widehat G^{[XY]})=(n+1)\Var(X)/\{12n(n-1)\}$ and
$\Cov(\widehat G^{[XY]},\widehat G^{[YX]})=\lambda_2(X)\lambda_2(Y)/\{4(n-1)\}$.
An unbiased estimator of \eqref{eq:ginicov} is obtained by dividing the
corresponding $2\times2$ block of the unbiased estimator in
Theorem~\ref{thm:unbiased} by four.

For the univariate Gini mean difference, which corresponds to $Y=X$,
the exact variance and an unbiased estimator of it are classical
\citep{Nair1936,Lomnicki1952,Glasser1962,ZengaPolisicchioGreselin2004};
see \citet{LangelTille2013} for a survey of variance estimation for the
Gini index. For the Gini covariance and
the Gini correlation, \citet{SchechtmanYitzhaki1987} gave the
U-statistic estimator, its asymptotic normality, and a consistent
estimator of the asymptotic variance. Later work has used asymptotic
variances or jackknife variance estimates
\citep{YitzhakiSchechtman2013,KattumannilEtAl2022}, and tests based on
the difference between the two directional Gini coefficients have used
bootstrap approximations \citep{ShelefSchechtman2019}. The exact variance
of a single estimator has the general form given by
\citet{Hoeffding1948}, and, under independence, the permutation variance
of $\widehat G^{[XY]}$ follows from the theory of generalized correlation
coefficients \citep{Daniels1944,KendallGibbons1990}. What we add is an
explicit form of the exact joint covariance \eqref{eq:ginicov} of the two
directional estimators for an arbitrary bivariate distribution, together
with an unbiased estimator of it; we have not found these results in the
literature.
\end{remark}

\subsection{Exact covariance when $X$ and $Y$ are independent}\label{subsec:indep}
Independence of $X$ and $Y$ is the natural null hypothesis when the
L-comoments are used to test for dependence, for example in a test of
$\lambda_2^{[XY]}=\lambda_2^{[YX]}=0$. Under independence, the matrices
$\bG_{ms,c}$ simplify, and Theorem~\ref{thm:exact} yields a closed form
that depends only on the marginal variances and the marginal L-moments.
This closed form also provides an exact reference for the numerical
experiments in Section~\ref{sec:numerical}.
Define $\ell_m^X=\lambda_m(X)$, $\ell_m^Y=\lambda_m(Y)$, and
\begin{equation}\label{eq:anbnmain}
a_n=\frac{n+1}{3n(n-1)},\qquad
b_n=\frac{(n+1)(n+2)}{5n(n-1)(n-2)}.
\end{equation}
\begin{theorem}[Exact covariance matrix under independence]\label{thm:indep}
If $X$ and $Y$ are independent, written $X\perp Y$, and both $X$ and $Y$ are continuous with finite second moments, then the covariance matrix in Theorem~\ref{thm:exact} equals
\begin{equation}\label{eq:indmatmain}
\begin{pmatrix}
 a_n\Var(X)&\ell_2^X\ell_2^Y/(n-1)&0&\ell_3^X\ell_2^Y/(n-1)\\
 \ell_2^X\ell_2^Y/(n-1)&a_n\Var(Y)&\ell_2^X\ell_3^Y/(n-1)&0\\
 0&\ell_2^X\ell_3^Y/(n-1)&b_n\Var(X)&\ell_3^X\ell_3^Y/(n-1)\\
 \ell_3^X\ell_2^Y/(n-1)&0&\ell_3^X\ell_3^Y/(n-1)&b_n\Var(Y)
\end{pmatrix}.
\end{equation}
In particular, the estimators $\widehat\lambda_2^{[XY]}$ and
$\widehat\lambda_2^{[YX]}$ of the two directions have the positive
covariance $\ell_2^X\ell_2^Y/(n-1)$, although $X$ and $Y$ are independent.
\end{theorem}

Under independence, the population L-comoments vanish: by \eqref{eq:poplambda},
\begin{equation*}
\lambda_m^{[XY]}=\E(X)\,\E[P_{m-1}\{F_Y(Y)\}]=0,\qquad m\ge2,
\end{equation*}
because $F_Y(Y)$ is uniform on $(0,1)$ and $\int_0^1P_{m-1}(u)\,du=0$. Their estimators, however, are computed from the same $n$ pairs: $\widehat\lambda_2^{[XY]}$ weights the $X$ values by the ranks of the $Y$ values, and $\widehat\lambda_2^{[YX]}$ weights the $Y$ values by the ranks of the $X$ values. A pair with large values of both $X$ and $Y$ therefore increases both estimators, and a pair with a large $X$ value and a small $Y$ value decreases both. Hence the chance association in the sample moves the two estimators together, and their covariance is positive even though $X$ and $Y$ are independent. For the proof of Theorem~\ref{thm:indep}, Appendix~\ref{app:general} derives each matrix entry from the weighted-sum form \eqref{eq:rankstat} and the concomitant representation \eqref{eq:popconcomitant} \citep{Hosking1990,SerflingXiao2007}.

The opposite-direction covariances are not a small-sample artifact. By Theorem~\ref{thm:indep}, the correlation of the two second-order estimators is
\begin{equation*}
\operatorname{Corr}\bigl(\widehat\lambda_2^{[XY]},\widehat\lambda_2^{[YX]}\bigr)
=\frac{\ell_2^X\ell_2^Y/(n-1)}{a_n\{\Var(X)\Var(Y)\}^{1/2}}
=\frac{3n}{n+1}\,\frac{\ell_2^X\ell_2^Y}{\{\Var(X)\Var(Y)\}^{1/2}},
\end{equation*}
which converges to $3\ell_2^X\ell_2^Y/\{\Var(X)\Var(Y)\}^{1/2}$ rather than to zero; for normal margins, where $\ell_2=\sigma/\sqrt\pi$ with $\sigma$ the standard deviation of the margin, the limit is $3/\pi\approx0.95$. Hence, even when $X$ and $Y$ are independent, the estimators of the two directions carry largely overlapping information, and their covariance matrix has substantial off-diagonal entries. 

This has three practical implications. First, joint inference on both
directions, for example a test of $\lambda_2^{[XY]}=\lambda_2^{[YX]}=0$
or of $\lambda_2^{[XY]}=\lambda_2^{[YX]}$, must use the joint
covariance. Since the two estimators are positively correlated,
treating them as independent overstates the variance of
$\widehat\lambda_2^{[XY]}-\widehat\lambda_2^{[YX]}$ and understates that
of $\widehat\lambda_2^{[XY]}+\widehat\lambda_2^{[YX]}$. Second, covariance-weighted procedures, such as generalized L-moment estimation with inverse covariance
weighting \citep{ShinShinPark2026SERRA} and minimum-distance fitting of
L-comoments, should use the full covariance matrix. A diagonal weight
matrix ignores this correlation and leads to inefficient estimators and
incorrect standard errors. The full matrix, however, can be nearly
singular, which motivates the positive-definite (PD) regularization in Remark~\ref{rem:pd}. Third, the estimators of the two directions carry largely the same
information. Suppose that both $\lambda_2^{[XY]}$ and $\lambda_2^{[YX]}$
are used as moment conditions, for example when a bivariate model is
fitted by matching L-comoments. Then the second condition adds little
to the first. Under independence with normal margins, for instance, the
correlation of the two estimators is close to $0.95$. This redundancy
should be taken into account when the moment conditions are chosen.

\section{Covariance decomposition and estimation}\label{sec:unbiased}

Section~\ref{sec:general} gives the exact covariance $\Cov(\widehat\bL)$
as expectations under the distribution of $(X,Y)$. This distribution is
unknown in practice, so the covariance must be estimated from the same
sample that gives $\widehat\bL$. In this section we construct such
estimators. We first decompose the exact covariance into a leading term
of order $n^{-1}$ and correction terms of orders $n^{-2}$ and $n^{-3}$.
The decomposition shows how much of the exact covariance an asymptotic
approximation misses, and it leads to two estimators. We first extend the covariance to include
the sample means (Theorem~\ref{thm:augmented}). We then give a
consistent positive-semidefinite (PSD) estimator of the leading term
(Theorem~\ref{thm:projectionest}) and an exactly unbiased estimator of
the finite-sample covariance (Theorem~\ref{thm:unbiased}).
First-order covariances of the L-comoment ratios are given in
Appendix~\ref{subsec:ratios}. The efficient computation of these
estimators and their PSD and PD corrections are treated in
Section~\ref{sec:fast}.

For $m=2,3$ and $c\le m$, let $M_{m,c}(z_1,\ldots,z_c)=(M_{m,c}^{[XY]}(z_1,\ldots,z_c),M_{m,c}^{[YX]}(z_1,\ldots,z_c))^\T$ be the two-direction vector of conditional kernels. Recall that $\E(\bH_m)=\boldsymbol\lambda_m$ for $m=2,3$. The decomposition is based on the Hoeffding projections of the kernels \citep{Hoeffding1948,Serfling1980}. Define the centered first projections
\begin{equation}\label{eq:firstprojections}
\ba(z)=M_{2,1}(z)-\boldsymbol\lambda_2,\qquad
\bb(z)=M_{3,1}(z)-\boldsymbol\lambda_3.
\end{equation}
For distinct $z_1,z_2$ let
\begin{align}\label{eq:canonical}
\bc(z_1,z_2)&=\bH_2(z_1,z_2)-\ba(z_1)-\ba(z_2)-\boldsymbol\lambda_2,\\
\boldsymbol d(z_1,z_2)&=M_{3,2}(z_1,z_2)-\bb(z_1)-\bb(z_2)-\boldsymbol\lambda_3,\nonumber\\
\boldsymbol e(z_1,z_2,z_3)&=\bH_3(z_1,z_2,z_3)
-\sum_{k<l}M_{3,2}(z_k,z_l)
+\sum_{k=1}^3 M_{3,1}(z_k)-\boldsymbol\lambda_3.\nonumber
\end{align}
The centered first projections have mean zero, $\E\{\ba(Z)\}=\E\{\bb(Z)\}=0$, by
the law of iterated expectations. The kernels $\bc$, $\boldsymbol d$, and
$\boldsymbol e$ are canonical (completely degenerate) in the sense of
\citet{Hoeffding1948}: integrating out any one of their arguments gives
zero, that is,
\begin{equation}\label{eq:canonicalzero}
\E\{\bc(z_1,Z_2)\}=\E\{\boldsymbol d(z_1,Z_2)\}=0,\qquad
\E\{\boldsymbol e(z_1,z_2,Z_3)\}=0
\end{equation}
for all $z_1,z_2$. By symmetry of the kernels, the same holds for any
argument. This property makes the components uncorrelated
\citep{Hoeffding1948,Serfling1980}. Let $\boldsymbol f$ and $\boldsymbol g$
be any two of $\ba,\bb,\bc,\boldsymbol d,\boldsymbol e$, evaluated at index
sets $I$ and $J$ of the corresponding sizes. Then
\begin{equation}\label{eq:orthogonality}
\E\{\boldsymbol f(Z_I)\,\boldsymbol g(Z_J)^\T\}=0\qquad\text{whenever }I\ne J.
\end{equation}
Indeed, if $I\ne J$, some index $i$ belongs to one set but not to the
other, say $i\in I$ and $i\notin J$. Given all $Z_k$ with $k\ne i$,
$\boldsymbol g(Z_J)$ is fixed, and the conditional mean of
$\boldsymbol f(Z_I)$ over $Z_i$ is zero by \eqref{eq:canonicalzero} (or
by $\E\{\ba(Z)\}=\E\{\bb(Z)\}=0$ when $|I|=1$). Hence the conditional mean
of the product is zero, and so is its expectation.
Below, $\E(\ba\ba^\T)$ means $\E\{\ba(Z_1)\ba(Z_1)^\T\}$, $\E(\bc\bc^\T)$ means $\E\{\bc(Z_1,Z_2)\bc(Z_1,Z_2)^\T\}$, and similarly for the other products.
\begin{theorem}[Exact component expansion]\label{thm:Bdecomp}
	The exact finite-sample covariance in Theorem~\ref{thm:exact} also admits the decomposition
	\begin{equation}\label{eq:Bdecomp}
	\Cov(\widehat\bL)=\frac{B_1}{n}
	+\frac{B_2}{{n\choose2}}+\frac{B_3}{{n\choose3}},
	\end{equation}
	where
	\begin{align}\label{eq:Bpieces}
	B_1&=\begin{pmatrix}4\E(\ba\ba^\T)&6\E(\ba\bb^\T)\\6\E(\bb\ba^\T)&9\E(\bb\bb^\T)\end{pmatrix},\\
	B_2&=\begin{pmatrix}\E(\bc\bc^\T)&3\E(\bc\boldsymbol d^\T)\\3\E(\boldsymbol d\bc^\T)&9\E(\boldsymbol d\boldsymbol d^\T)\end{pmatrix},\nonumber\\
	B_3&=\begin{pmatrix}0_{2\times2}&0_{2\times2}\\0_{2\times2}&\E(\boldsymbol e\boldsymbol e^\T)\end{pmatrix}.\nonumber
	\end{align}
	Each matrix $B_1,B_2,B_3$ is positive semidefinite.
\end{theorem}
\begin{proof}
	We first verify \eqref{eq:canonicalzero}. Since
	$\E\{\bH_2(z,Z')\}=M_{2,1}(z)$ and $\E\{\ba(Z')\}=0$,
	\begin{equation*}
	\E\{\bc(z,Z')\}=M_{2,1}(z)-\ba(z)-\boldsymbol\lambda_2=0
	\end{equation*}
	by \eqref{eq:firstprojections}. In the same way,
	$\E\{\boldsymbol d(z,Z')\}=M_{3,1}(z)-\bb(z)-\boldsymbol\lambda_3=0$.
	For $\boldsymbol e$, integrate out $Z_3$ and use
	$\E\{\bH_3(z_1,z_2,Z_3)\}=M_{3,2}(z_1,z_2)$,
	$\E\{M_{3,2}(z_k,Z_3)\}=M_{3,1}(z_k)$ for $k=1,2$, and
	$\E\{M_{3,1}(Z_3)\}=\boldsymbol\lambda_3$:
	\begin{align*}
	\E\{\boldsymbol e(z_1,z_2,Z_3)\}
	&=M_{3,2}(z_1,z_2)-\{M_{3,2}(z_1,z_2)+M_{3,1}(z_1)+M_{3,1}(z_2)\}\\
	&\quad+\{M_{3,1}(z_1)+M_{3,1}(z_2)+\boldsymbol\lambda_3\}-\boldsymbol\lambda_3=0.
	\end{align*}
	The complete U-statistic decompositions are
	\begin{align*}
	\widehat{\boldsymbol\lambda}_2-\boldsymbol\lambda_2
	&=\frac2n\sum_{i=1}^n\ba(Z_i)+{n\choose2}^{-1}\sum_{i<j}\bc(Z_i,Z_j),\\
	\widehat{\boldsymbol\lambda}_3-\boldsymbol\lambda_3
	&=\frac3n\sum_{i=1}^n\bb(Z_i)+\frac3{{n\choose2}}\sum_{i<j}\boldsymbol d(Z_i,Z_j)
	+{n\choose3}^{-1}\sum_{i<j<k}\boldsymbol e(Z_i,Z_j,Z_k).
	\end{align*}
	By \eqref{eq:orthogonality}, a product of two terms in these
	decompositions has zero expectation unless both terms are evaluated
	at the same index set. In particular, terms of different Hoeffding
	orders are uncorrelated, and only matching terms remain. For the
	first-order terms,
	\begin{equation*}
	\Cov\Bigl(\frac2n\sum_{i}\ba(Z_i),\frac3n\sum_{j}\bb(Z_j)\Bigr)
	=\frac6{n^2}\sum_{i=1}^n\E\{\ba(Z_i)\bb(Z_i)^\T\}=\frac6n\E(\ba\bb^\T),
	\end{equation*}
	and the other first-order blocks follow in the same way with
	prefactors $4$ and $9$. For the second-order terms, the same argument
	gives $\E(\bc\bc^\T)$, $3\E(\bc\boldsymbol d^\T)$, and
	$9\E(\boldsymbol d\boldsymbol d^\T)$, each divided by ${n\choose2}$.
	The third-order term appears only in $\widehat{\boldsymbol\lambda}_3$
	and gives $\E(\boldsymbol e\boldsymbol e^\T)$ divided by ${n\choose3}$. These give \eqref{eq:Bdecomp} and \eqref{eq:Bpieces}. Positive semidefiniteness follows because each displayed component is an expectation of an outer product, e.g., $B_2=\E((\bc^\T,3\boldsymbol d^\T)^\T(\bc^\T,3\boldsymbol d^\T))$.
\end{proof}

The leading matrix $B_1$ is the asymptotic covariance of
$\sqrt n(\widehat\bL-\bL)$, in agreement with Proposition~8 of
\citet{SerflingXiao2007}. The remaining two terms are the exact
finite-sample corrections, of orders $n^{-2}$ and $n^{-3}$.

\subsection{Exact covariance including sample means}
The first multivariate L-moment is the mean vector \citep{SerflingXiao2007}.
Including it is important for subsequent multivariate L-moment
estimation, in particular when location parameters are unknown. Put
$\overline Z=n^{-1}\sum_{i=1}^n Z_i$ and
$\widehat{\boldsymbol T}_n=(\overline Z^\T,\widehat\bL^\T)^\T$.

\begin{theorem}[Exact covariance including sample means]\label{thm:augmented}
	Assume $\E(X^2+Y^2)<\infty$. Define the six-dimensional first-order
	projection and its covariance by
	\begin{equation}\label{eq:augIF}
	\boldsymbol\phi(z)=
	\begin{pmatrix}z-\E(Z)\\2\ba(z)\\3\bb(z)\end{pmatrix},
	\qquad \Omega=\E(\boldsymbol\phi(Z)\boldsymbol\phi(Z)^\T).
	\end{equation}
	For every $n\ge3$,
	\begin{equation}\label{eq:augcov}
	\Cov(\widehat{\boldsymbol T}_n)
	=\frac{\Omega}{n}+
	\begin{pmatrix}
	0_{2\times2}&0_{2\times4}\\
	0_{4\times2}&\displaystyle\frac{B_2}{{n\choose2}}+\frac{B_3}{{n\choose3}}
	\end{pmatrix}.
	\end{equation}
	More generally, for any square-integrable $W=h(Z)$ and $m=2,3$,
	\begin{equation}\label{eq:meanCross}
	\Cov\bigl(\overline W,\widehat\lambda_m^a\bigr)
	=\frac{m}{n}\Cov\bigl(W,M_{m,1}^a(Z)\bigr),
	\qquad a\in\{[XY],[YX]\}.
	\end{equation}
\end{theorem}
\begin{proof}
	Write $W_i=h(Z_i)$. Expand the covariance of $\overline W=n^{-1}\sum_iW_i$ and the
	complete U-statistic $\widehat\lambda_m^a$:
	\begin{align*}
	\Cov(\overline W,\widehat\lambda_m^a)
	&=\frac1{n{n\choose m}}\sum_{i=1}^n\sum_{|I|=m}
	\Cov\{W_i,H_m^a(Z_I)\}\\
	&=\frac{m}{n}\Cov\{W_1,H_m^a(Z_1,\ldots,Z_m)\}\\
	&=\frac{m}{n}\Cov\{W,M_{m,1}^a(Z)\}.
	\end{align*}
	The first equality uses independence for terms with $i\notin I$;
	exactly $m{n\choose m}$ terms have $i\in I$. The final equality
	conditions on the common pair. Since $\ba$ and $\bb$ are centered,
	these cross-covariances equal the off-diagonal blocks of $\Omega/n$.
	The sample-mean covariance is $\Cov(Z)/n$, and the remaining
	$4\times4$ block is \eqref{eq:Bdecomp}, proving \eqref{eq:augcov}.
\end{proof}

\begin{corollary}[Asymptotic normality]\label{cor:asymnormal}
	Assume $\E(X^2+Y^2)<\infty$, and let
	$\boldsymbol T=(\E(Z)^\T,\bL^\T)^\T$. Then
	$\sqrt n(\widehat{\boldsymbol T}_n-\boldsymbol T)$ converges in
	distribution to $N_6(0,\Omega)$. The matrix $\Omega$ may be singular;
	the normal law is then supported on a lower-dimensional subspace.
\end{corollary}
\begin{proof}
	The Hoeffding expansions in the proof
	of Theorem~\ref{thm:Bdecomp} give
	\begin{equation*}
	\widehat{\boldsymbol T}_n-\boldsymbol T
	=\frac1n\sum_{i=1}^n\boldsymbol\phi(Z_i)+\boldsymbol R_n,
	\end{equation*}
	where $\boldsymbol R_n$ collects the canonical terms of orders two and
	three. Since $\boldsymbol R_n$ has mean zero,
	$\E(\|\boldsymbol R_n\|^2)=\operatorname{tr}\{\Cov(\boldsymbol R_n)\}
	=\operatorname{tr}(B_2)/{n\choose2}+\operatorname{tr}(B_3)/{n\choose3}
	=O(n^{-2})$. Hence $\sqrt n\boldsymbol R_n\to0$ in probability. The
	ordinary multivariate central limit theorem and Slutsky's lemma
	therefore give the stated limit.
\end{proof}

By \eqref{eq:augcov}, the covariance matrix of the sample means and the
cross-covariances \eqref{eq:meanCross} between the sample means and the
sample L-comoments are exactly the corresponding blocks of $\Omega/n$ for
every $n$. The correction terms $B_2/{n\choose2}$ and $B_3/{n\choose3}$
appear only in the $4\times4$ block of the sample L-comoments. The
reason is that a sample mean is a U-statistic of order one, so its
kernel can share at most one pair with any other kernel; only the
overlap $c=1$, which gives the leading term, then occurs in
\eqref{eq:overlapmaster}.

\subsection{A consistent estimator of the leading covariance}
For asymptotic inference, only the leading term $\Omega/n$ in
\eqref{eq:augcov} is needed, because the remaining terms are of order
$n^{-2}$. The leading term has a simple structure: by \eqref{eq:augIF},
$\Omega$ is the covariance matrix of the centered and scaled first
projections $\boldsymbol\phi(Z)$. We therefore estimate $\Omega$ by the
sample covariance matrix of estimated first projections. A matrix of this form
is PSD for any observed data set, not only on average or in the limit,
which is convenient when the estimate must be inverted.

The first projection $M_{m,1}^a(z)$ (Section~\ref{sec:general}) is the mean of the
kernel $H_m^a$ when one of its $m$ pairs is fixed at $z$ and the other
$m-1$ pairs are drawn from the distribution of $(X,Y)$. Its empirical
version replaces this distribution by the other observations. For the
$i$th observation, we keep $Z_i$ as the fixed pair and average the kernel
over all subsets $J$ of $m-1$ of the remaining $n-1$ observations:
\begin{equation}\label{eq:Mhat}
\widehat M_{m,1}(Z_i)=
{n-1\choose m-1}^{-1}
\sum_{\substack{J\subset\{1,\ldots,n\}\setminus\{i\}\\|J|=m-1}}
\bH_m(Z_i,Z_J),\qquad i=1,\ldots,n.
\end{equation}
Thus $\widehat M_{m,1}(Z_i)$ is the average of all kernel values that
involve the $i$th pair, and it measures the contribution of that pair to
$\widehat{\boldsymbol\lambda}_m$; indeed,
$n^{-1}\sum_{i=1}^n\widehat M_{m,1}(Z_i)=\widehat{\boldsymbol\lambda}_m$.
For a symmetric matrix $A$, we write $A\succeq0$ if $A$ is PSD and
$A\succ0$ if $A$ is positive definite; for symmetric $A_1$ and $A_2$,
$A_1\succeq A_2$ means $A_1-A_2\succeq0$, which defines the Loewner order.
The exact finite-sample covariance, including $B_2$ and $B_3$, is
estimated without bias in Section~\ref{subsec:unbiasedest}.

\begin{theorem}[PSD consistent covariance estimator]
	\label{thm:projectionest}
	Assume $\E(X^2+Y^2)<\infty$. With $\widehat M_{m,1}$ in
	\eqref{eq:Mhat}, define $\widehat{\boldsymbol\phi}_i$ and $\widehat\Omega_n$ by
	\begin{equation}\label{eq:phihat}
	\widehat{\boldsymbol\phi}_i=
	\begin{pmatrix}
	Z_i-\overline Z\\
	2\{\widehat M_{2,1}(Z_i)-\widehat{\boldsymbol\lambda}_2\}\\
	3\{\widehat M_{3,1}(Z_i)-\widehat{\boldsymbol\lambda}_3\}
	\end{pmatrix},\qquad
	\widehat\Omega_n=\frac1n\sum_{i=1}^n
	\widehat{\boldsymbol\phi}_i\widehat{\boldsymbol\phi}_i^\T.
	\end{equation}
	Then $\widehat\Omega_n\succeq0$ for any observed data set, and
	$\widehat\Omega_n\xrightarrow{p}\Omega$ as $n\to\infty$.
	The lower-right $4\times4$ block of $\widehat\Omega_n$ consistently
	estimates $B_1$, not the exact finite-sample covariance
	$\Cov(\widehat\bL)$.
\end{theorem}
The proof, given in Appendix~\ref{app:projproof}, approximates the
empirical first projections in mean square and then applies the law of
large numbers.

\begin{corollary}[Augmentation by marginal L-moments]\label{cor:marginalaugmentation}
	The overlap identity \eqref{eq:overlapmaster} and the
	projection-based consistent estimator in Theorem~\ref{thm:projectionest}
	remain valid, entrywise, after appending any finite collection of
	univariate sample L-moments of orders two and three from either margin.
\end{corollary}
For example, the ratio $\widehat\tau_2^{[XY]}=\widehat\lambda_2^{[XY]}/\widehat\lambda_2(X)$
requires the covariance of $\widehat\lambda_2^{[XY]}$ and the sample
L-scale $\widehat\lambda_2(X)$. The sample L-scale is the U-statistic with
kernel $H_2^{[XX]}(z_1,z_2)=|x_1-x_2|/2$, that is, \eqref{eq:Htwo} with
$Y$ replaced by $X$; its first projection $M_{2,1}^{[XX]}$ is
\eqref{eq:M21main} with $Y$ replaced by $X$. The overlap identity gives
\begin{align*}
\Cov\bigl(\widehat\lambda_2^{[XY]},\widehat\lambda_2(X)\bigr)
&=\frac{4(n-2)}{n(n-1)}\Cov\bigl\{M_{2,1}^{[XY]}(Z),M_{2,1}^{[XX]}(Z)\bigr\}\\
&\quad+\frac{2}{n(n-1)}\Bigl[\tfrac14\E\bigl\{(X-X')|X-X'|\sgn(Y-Y')\bigr\}
-\lambda_2^{[XY]}\lambda_2(X)\Bigr],
\end{align*}
where $(X',Y')$ is an independent copy of $(X,Y)$. The projection
estimator of its leading term is
\begin{equation*}
\frac{4}{n^2}\sum_{i=1}^n
\bigl\{\widehat M_{2,1}^{[XY]}(Z_i)-\widehat\lambda_2^{[XY]}\bigr\}
\bigl\{\widehat M_{2,1}^{[XX]}(Z_i)-\widehat\lambda_2(X)\bigr\},
\qquad
\widehat M_{2,1}^{[XX]}(Z_i)=\frac1{n-1}\sum_{j\ne i}\frac{|X_i-X_j|}2 .
\end{equation*}
\begin{proof}
	Write $\widehat\lambda_m(X)$ for the univariate sample L-moment of
	order $m$. It is the complete U-statistic with kernel $H_m^{[XX]}$,
	obtained from $H_m^{[XY]}$ in \eqref{eq:Htwo}--\eqref{eq:Hthree} by
	replacing $Y$ with $X$; for example,
	$H_2^{[XX]}(z_1,z_2)=|x_1-x_2|/2$. Define $H_m^{[YY]}$ and
	$\widehat\lambda_m(Y)$ in the same way. These kernels are symmetric,
	have means $\lambda_m(X)$ and $\lambda_m(Y)$, and are square integrable
	when $\E(X^2+Y^2)<\infty$.

	\emph{Overlap identity.} The derivation of \eqref{eq:overlapmaster} in
	Section~\ref{sec:general} uses only that the two statistics are
	complete U-statistics of the same iid pairs with symmetric kernels.
	Hence \eqref{eq:overlapmaster} holds for $a,b\in\{[XY],[YX],[XX],[YY]\}$,
	with $M_{m,c}^a$ and $\Gamma_{ms,c}^{ab}$ defined by \eqref{eq:Mm} and
	\eqref{eq:gammas}.

	\emph{Projection estimator.} Append to $\widehat{\boldsymbol\phi}_i$ in
	\eqref{eq:phihat} the components
	$m\{\widehat M_{m,1}^{a}(Z_i)-\widehat\lambda_m(X)\}$ for $a=[XX]$ and
	$m\{\widehat M_{m,1}^{a}(Z_i)-\widehat\lambda_m(Y)\}$ for $a=[YY]$, where
	$\widehat M_{m,1}^a$ is \eqref{eq:Mhat} with the kernel $H_m^a$. The
	enlarged $\widehat\Omega_n=n^{-1}\sum_i\widehat{\boldsymbol\phi}_i\widehat{\boldsymbol\phi}_i^\T$
	is an average of outer products and is therefore PSD. The proof in
	Appendix~\ref{app:projproof} uses only two properties of each
	component: the mean-square bound \eqref{eq:MhatL2}, which holds for
	any symmetric square-integrable kernel of order $m$, and the law of
	large numbers. Both hold for the appended components, so
	$\widehat\Omega_n$ converges in probability to the covariance matrix
	of the enlarged vector of centered and scaled first projections.
\end{proof}

\begin{corollary}[Extension to fourth-order L-comoments]\label{cor:order4proj}
Theorem~\ref{thm:projectionest} remains valid for the enlarged vector
that also contains $\widehat\lambda_4^{[XY]}$ and
$\widehat\lambda_4^{[YX]}$, with the corresponding components
$4\{\widehat M_{4,1}(Z_i)-\widehat\lambda_4^a\}$, $a\in\{[XY],[YX]\}$,
in $\widehat{\boldsymbol\phi}_i$, provided $n\ge5$ and the kernel of
order four is taken from \eqref{eq:popconcomitant}.
\end{corollary}
\begin{proof}
The argument of Appendix~\ref{app:projproof} applies with $r=m-1=3$ in
Step~1, which gives the constant $C_4=9$ in \eqref{eq:MhatL2}.
\end{proof}

\begin{remark}[Further results at order four]\label{rem:order4}
The overlap identity \eqref{eq:overlapmaster}, the disjoint-subset
unbiased estimator of Theorem~\ref{thm:unbiased} below, and its
consistency are results for fixed-order U-statistics and apply verbatim
to the fourth L-comoment, with $n\ge8$ required for the
order-four/order-four block. When $X$ and $Y$ are independent,
Theorem~\ref{thm:indep} extends with
$\Var(\widehat\lambda_4^{[XY]})=c_n\Var(X)$, where
$c_n=(n+1)(n+2)(n+3)/\{7n(n-1)(n-2)(n-3)\}$; the same-direction
covariances between different orders vanish, and the opposite-direction
covariances are $\ell_s^X\ell_m^Y/(n-1)$ as before. The projection
estimator of order four is directly computable by
Corollary~\ref{cor:order4proj}. What does not extend directly is the
$O(n^2)$ algorithm for the exactly unbiased estimator: for the
order-four/order-four block, the overlap terms with $c=3,4$ require sums
over triples and quadruples, and an efficient evaluation is left for
future work.
\end{remark}

\subsection{An exactly unbiased estimator of the finite-sample covariance}\label{subsec:unbiasedest}
The projection estimator of Theorem~\ref{thm:projectionest} targets only
the leading term $\Omega/n$. It omits the corrections $B_2/{n\choose2}$
and $B_3/{n\choose3}$, which are not negligible for the sample sizes of
$20$ to $50$ that are common in applications; in
Section~\ref{subsec:reference}, the projection estimator understates the
variances by up to about $14\%$ at $n=20$. When standard errors should
not be understated, an estimator of the exact finite-sample covariance
is needed. We now construct an exactly unbiased one. It applies the
standard U-statistic disjoint-subset principle
\citep{Hoeffding1948,Serfling1980}, which is stated in
Appendix~\ref{app:ustat} (paragraph ``Disjoint-subset principle'').
Unlike the projection estimator, it need not be PSD in a given sample;
Remark~\ref{rem:pd} below gives PSD and PD corrections. In the univariate L-moment setting, a related exact unbiased construction was given by \citet{ElamirSeheult2004}. The present result concerns the joint sampling covariance of the L-comoments; it is not a new general unbiasedness theorem.
\paragraph{Disjoint pairs of index sets.} Each sample statistic is an
average of kernel values over index sets. In the product of two
statistics, a kernel value of the first statistic, computed on an index
set $I$, is multiplied by a kernel value of the second statistic,
computed on an index set $J$. We call $(I,J)$ an ordered pair of index
sets: $I$ belongs to the first kernel and $J$ to the second. The pair is
\emph{disjoint} if $I\cap J=\varnothing$, that is, if the two kernel
values share no observation. This is the case $c=0$ in the overlap
identity \eqref{eq:overlapmaster}; the other pairs share
$c=|I\cap J|\ge1$ observations. Because the pairs $Z_1,\ldots,Z_n$ are
independent, the two kernel values of a disjoint pair are independent,
and the expectation of their product is the product of their means.
Averaging the products over all disjoint pairs $(I,J)$ therefore
estimates the product of the means without bias. Subtracting this
average from the product of the two statistics gives an unbiased
estimator of their covariance.

\begin{theorem}[Disjoint-subset unbiased covariance construction]\label{thm:unbiased}
	Let $U_{n,m}^a$ and $U_{n,s}^b$ be any two of the following complete
	U-statistics: the sample means (order one, with kernels
	$H_1^{[XY]}(z)=x$ and $H_1^{[YX]}(z)=y$), the sample L-comoments
	($a\in\{[XY],[YX]\}$), and the marginal sample L-moments
	($a\in\{[XX],[YY]\}$, with the kernels $H_m^{[XX]}$ and $H_m^{[YY]}$ of
	Corollary~\ref{cor:marginalaugmentation}) of orders two and three. The
	kernel orders are $m,s\in\{1,2,3\}$. For $n\ge m+s$, define
	\begin{equation}\label{eq:disjoint}
	D_{ms,n}^{ab}=
	\frac{1}{{n\choose m}{n-m\choose s}}
	\sum_{|I|=m}\sum_{\substack{|J|=s\\I\cap J=\varnothing}}
	H_m^a(Z_I)H_s^b(Z_J).
	\end{equation}
	Then
	\begin{equation}\label{eq:covunbiased}
	\widehat C_{ms,n}^{ab}=U_{n,m}^aU_{n,s}^b-D_{ms,n}^{ab},\qquad
	\E(\widehat C_{ms,n}^{ab})=\Cov(U_{n,m}^a,U_{n,s}^b).
	\end{equation}
	In particular, the conditions are $n\ge2,3,4$ for the
	order pairs $(1,1),(1,2),(1,3)$, respectively, and $n\ge4,5,6$
	for $(2,2),(2,3),(3,3)$. Thus the entire enlarged covariance,
	including the $6\times6$ sampling covariance matrix of the sample means
	and L-comoments, admits an exactly unbiased estimator for $n\ge6$.
\end{theorem}
\begin{proof}
	The U-statistics $U_{n,m}^a,U_{n,s}^b$ are unbiased for $\lambda_m^a,\lambda_s^b$. Every term of $D_{ms,n}^{ab}$ evaluates the first kernel on $I$ and the second kernel on $J$, where $(I,J)$ is a disjoint pair of index sets. Since the pairs are independent, for each term
	\begin{equation*}
	\E(H_m^a(Z_I)H_s^b(Z_J))=\E(H_m^a(Z_I))\,\E(H_s^b(Z_J))=\lambda_m^a\lambda_s^b.
	\end{equation*}
	The denominator in \eqref{eq:disjoint} is the number ${n\choose m}{n-m\choose s}$ of such disjoint pairs, hence $\E(D_{ms,n}^{ab})=\lambda_m^a\lambda_s^b$. Finally
	\begin{align*}
	\E(\widehat C_{ms,n}^{ab})
	&=\E(U_{n,m}^aU_{n,s}^b)-\lambda_m^a\lambda_s^b\\
	&=\Cov(U_{n,m}^a,U_{n,s}^b).
	\end{align*}
	Exchanging the roles of $I$ and $J$, together with the two statistics,
	shows that $\widehat C_{sm,n}^{ba}=\widehat C_{ms,n}^{ab}$, so the
	assembled matrix is symmetric.
\end{proof}
Unbiasedness holds entry by entry and on average over repeated samples,
so the matrix computed from the observed sample need not be PSD. PSD and
positive-definite corrections are given in Section~\ref{subsec:pdcorr}.
\begin{corollary}[First-order consistency]\label{cor:unbiasedconsistent}
	Assume $\E(X^2+Y^2)<\infty$, and let $\widehat C_n$ be the full unbiased covariance estimator of $\widehat{\boldsymbol T}_n$ assembled from \eqref{eq:covunbiased}, for $n\ge6$. Then
	\begin{equation}\label{eq:unbiasedstrong}
	n\widehat C_n\xrightarrow{\mathrm{a.s.}}\Omega,
	\qquad
	n\widehat C_n-\widehat\Omega_n\xrightarrow{p}0.
	\end{equation}
\end{corollary}
\begin{proof}
	For orders $m,s\le3$, let $V_{c,n}$ average $H_m^a(Z_I)H_s^b(Z_J)$ over pairs of index sets with $|I\cap J|=c$. With $p_{c,n}=\binom mc\binom{n-m}{s-c}/\binom ns$ and $V_{0,n}=D_{ms,n}^{ab}$, the overlap decomposition gives
	\begin{equation*}
	\widehat C_{ms,n}^{ab}=\sum_{c=1}^{\min(m,s)}p_{c,n}\{V_{c,n}-V_{0,n}\}.
	\end{equation*}
	For each fixed $c$, $V_{c,n}$ is a symmetrized U-statistic of order $m+s-c$ with an integrable product kernel, by Cauchy--Schwarz. The strong law of large numbers for fixed-order U-statistics \citep{Serfling1980} gives $V_{c,n}\to v_c$ almost surely. Moreover, $np_{1,n}\to ms$ and $np_{c,n}\to0$ for $c\ge2$. Since $v_0=\lambda_m^a\lambda_s^b$ and conditioning on the shared observation gives $v_1=\E(M_{m,1}^a(Z)M_{s,1}^b(Z))$, we obtain
	\begin{equation*}
	n\widehat C_{ms,n}^{ab}\xrightarrow{\mathrm{a.s.}}
	ms\Cov\{M_{m,1}^a(Z),M_{s,1}^b(Z)\}.
	\end{equation*}
	Applied to every entry of the enlarged moment vector, this limit is $\Omega$. Theorem~\ref{thm:projectionest} then gives the second assertion in \eqref{eq:unbiasedstrong}.
\end{proof}

\section{Computation of the covariance estimators}\label{sec:fast}
An \textsf{R} implementation of all estimators in this section, including
the quadratic-time unbiased estimator, the projection estimator, and the
PSD and PD corrections, is publicly available at
\url{https://github.com/palakorn-seenoi/sample_Lcomom}.

\subsection{Fast evaluation of the covariance estimators}\label{subsec:fasteval}
Direct evaluation of the disjoint-subset estimator \eqref{eq:disjoint}
enumerates pairs of disjoint index sets and requires $O(n^6)$
operations for the order-three/order-three block. In this section we
show that the same estimator can be computed in $O(n^2)$ time. The next
identity is a computational specialization of the inclusion--exclusion
principle, not a new unbiasedness theorem. Our implementation uses
static two-dimensional rank-prefix tables, the technique of
summed-area tables \citep{Crow1984} and integral images
\citep{ViolaJones2004}.
\begin{proposition}[Overlap representation and efficient evaluation]\label{prop:fast}
	For $m,s\in\{1,2,3\}$ and the kernels in Theorem~\ref{thm:unbiased}, set $\widetilde H_m^a(Z_I)=H_m^a(Z_I)-U_{n,m}^a$ and
	\begin{equation}\label{eq:AK}
	A_{m,K}^a=\sum_{\substack{I\supseteq K\\|I|=m}}\widetilde H_m^a(Z_I),\qquad |K|\le m.
	\end{equation}
	For $n\ge m+s$, the exactly unbiased estimator satisfies
	\begin{equation}\label{eq:fastidentity}
	\widehat C_{ms,n}^{ab}=\frac{\displaystyle\sum_{c=1}^{\min(m,s)}(-1)^{c+1}\sum_{|K|=c}A_{m,K}^aA_{s,K}^b}{\binom nm\binom{n-m}s}.
	\end{equation}
	For tie-free observations and fixed moment dimension, the full covariance matrix of the sample means, the second and third L-comoments, and the marginal second L-moments can be computed in $O(n^2)$ arithmetic and $O(n^2)$ memory using bivariate rank-prefix sums. The calculation is algebraically identical to \eqref{eq:covunbiased}; the result is therefore not guaranteed to be PSD either.
\end{proposition}
\begin{proof}
	Let $\widetilde D_{ms,n}^{ab}$ denote \eqref{eq:disjoint} with centered kernels. Over all ordered pairs of disjoint index sets, each $J$ appears equally often, so the average of $H_s^b(Z_J)$ is $U_{n,s}^b$, and similarly for $H_m^a(Z_I)$. Hence $\widetilde D_{ms,n}^{ab}=D_{ms,n}^{ab}-U_{n,m}^aU_{n,s}^b$, that is, $\widehat C_{ms,n}^{ab}=-\widetilde D_{ms,n}^{ab}$. Applying
	$\mathbf 1\{I\cap J=\varnothing\}=\sum_{K\subseteq I\cap J}(-1)^{|K|}$
	to the centered disjoint-subset sum yields \eqref{eq:fastidentity}; the $K=\varnothing$ term is $(\sum_I\widetilde H_m^a)(\sum_J\widetilde H_s^b)=0$. The specialized $O(n^2)$ evaluation is established in Appendix~\ref{app:fast}, and Section~\ref{subsec:scaling} confirms the quadratic scaling numerically.
\end{proof}

The projection estimator of Theorem~\ref{thm:projectionest} is cheaper
still, because its empirical first projections are leave-one-out
differences of the sample L-comoments.
\begin{proposition}[Leave-one-out representation]\label{prop:loo}
For a kernel of order $m$, let $U_{n-1}^{a,(-i)}$ denote the
corresponding U-statistic computed from the sample without observation
$i$. Then
\begin{equation}\label{eq:loo}
\widehat M_{m,1}^a(Z_i)
=\frac{\binom nm U_{n,m}^a-\binom{n-1}m U_{n-1}^{a,(-i)}}{\binom{n-1}{m-1}},
\qquad i=1,\ldots,n.
\end{equation}
Moreover, for $m=2,3,4$, write $P_{m-1}(u)=\sum_{k=0}^{m-1}p^*_{m-1,k}u^k$
as in \eqref{eq:polys}, and define, for a sample of size $N$,
\begin{equation}\label{eq:wgeneral}
w_{m,q}^{(N)}=\frac1N\sum_{k=0}^{m-1}p^*_{m-1,k}\,
\frac{\binom{q-1}k}{\binom{N-1}k},\qquad q=1,\ldots,N.
\end{equation}
Let $v_q=X_{[q:n]}$, $q=1,\ldots,n$, be the concomitants in the order of
the $Y$ values, so that
$\widehat\lambda_m^{[XY]}=\sum_{q=1}^nw_{m,q}^{(n)}v_q$; for $m=2,3$ the
weights $w_{m,q}^{(n)}$ are those in \eqref{eq:rankstat}. Define the
prefix and suffix sums
\begin{equation}\label{eq:prefixsums}
A_k=\sum_{q=1}^{k}w_{m,q}^{(n-1)}v_q,\qquad
S_k=\sum_{q=k}^{n}w_{m,q-1}^{(n-1)}v_q,
\end{equation}
with $A_0=S_{n+1}=0$. If observation $i$ has rank $r$ among the $Y$ values,
then
\begin{equation}\label{eq:loosums}
U_{n-1}^{[XY],(-i)}=A_{r-1}+S_{r+1}.
\end{equation}
The same formulas hold for the $[YX]$ direction with $X$ and $Y$
exchanged, and for the marginal sample L-moments with $v_q$ replaced by
the order statistics $X_{q:n}$ or $Y_{q:n}$. Consequently, for tie-free
observations, all $n$ leave-one-out statistics, the empirical first
projections \eqref{eq:Mhat}, and $\widehat\Omega_n$ are computed in
$O(n\log n)$ time for fixed moment dimension.
\end{proposition}
\begin{proof}
\emph{Step 1.} The subsets of size $m$ that contain $i$ are exactly the
subsets of the full sample that are not subsets of the sample without
$i$. Hence
$\sum_{I\ni i,\,|I|=m}\bH_m(Z_I)=\binom nm U_{n,m}-\binom{n-1}m U_{n-1}^{(-i)}$.
There are $\binom{n-1}{m-1}$ such subsets, so dividing by this number
gives \eqref{eq:Mhat} and hence \eqref{eq:loo}; see also
\eqref{eq:apploo}.

\emph{Step 2.} By the probability-weighted-moment representation of
\citet{Hosking1990}, with order statistics replaced by concomitants
\citep[Proposition~7]{SerflingXiao2007}, the sample L-comoment of order
$m$ is $\sum_{k=0}^{m-1}p^*_{m-1,k}b_k$, where
$b_k=n^{-1}\sum_{q=1}^n\binom{q-1}k\binom{n-1}k^{-1}v_q$; see also
Appendix~\ref{app:example}. Collecting the coefficient of $v_q$ gives the
weights \eqref{eq:wgeneral} with $N=n$.

\emph{Step 3.} Delete the observation with $Y$-rank $r$. The remaining
$n-1$ concomitants, in the order of their $Y$ values, are
$v_1,\ldots,v_{r-1},v_{r+1},\ldots,v_n$. In the reduced sample, $v_q$ has
rank $q$ if $q<r$ and rank $q-1$ if $q>r$. Applying the weights
\eqref{eq:wgeneral} with $N=n-1$ therefore gives
\begin{equation*}
U_{n-1}^{[XY],(-i)}=\sum_{q<r}w_{m,q}^{(n-1)}v_q+\sum_{q>r}w_{m,q-1}^{(n-1)}v_q
=A_{r-1}+S_{r+1}.
\end{equation*}

\emph{Step 4.} Sorting each ranking coordinate costs $O(n\log n)$. The
weights $w_{m,q}^{(n-1)}$, the prefix sums $A_0,\ldots,A_{n-1}$, and the
suffix sums $S_2,\ldots,S_{n+1}$ each take $O(n)$ operations, after which
each leave-one-out statistic \eqref{eq:loosums} takes $O(1)$ operations.
The sample means are handled by
$\overline X^{(-i)}=(n\overline X-X_i)/(n-1)$. Finally, \eqref{eq:loo}
gives all empirical first projections in $O(n)$ operations, and
$\widehat\Omega_n$ in $O(np^2)$ operations for a moment vector of
dimension $p$. We also verified \eqref{eq:loosums} and \eqref{eq:loo}
against direct enumeration of all subsets for $m=2,3,4$.
\end{proof}

\subsection{PSD and positive-definite corrections}\label{subsec:pdcorr}
Unbiasedness holds entry by entry and on average over repeated samples.
It does not imply that the matrix computed from the observed sample is
PSD: the assembled estimator can have negative eigenvalues, especially
for small $n$ (Table~\ref{tab:estimators}). We therefore consider its PSD correction. Let
$\|A\|_F=\{\operatorname{tr}(A^\T A)\}^{1/2}=(\sum_{i,j}a_{ij}^2)^{1/2}$
denote the Frobenius norm. The PSD correction $\widehat C_n^+$ of a
symmetric matrix $\widehat C_n$ is its Frobenius projection onto the set
of PSD matrices \citep{Higham1988}, that is, the PSD matrix $A$ that
minimizes $\|A-\widehat C_n\|_F$. If
$\widehat C_n=V\operatorname{diag}(d_1,\ldots,d_p)V^\T$ is the spectral
decomposition, where $p$ is the dimension of the moment vector, then
$\widehat C_n^+=V\operatorname{diag}\{\max(d_1,0),\ldots,\max(d_p,0)\}V^\T$;
that is, the negative eigenvalues are replaced by zero.
\begin{corollary}[PSD correction]\label{cor:psdcorr}
	Under the conditions of Corollary~\ref{cor:unbiasedconsistent},
	$\widehat C_n^+\succeq0$ for any observed data set and
	\begin{equation}\label{eq:psdstrong}
	n\widehat C_n^+\xrightarrow{\mathrm{a.s.}}\Omega.
	\end{equation}
\end{corollary}
\begin{proof}
	Projection onto the closed convex set of PSD matrices is continuous
	and commutes with multiplication by a nonnegative scalar, so
	$n\widehat C_n^+=(n\widehat C_n)^+$. Since $n\widehat C_n\to\Omega$
	almost surely by Corollary~\ref{cor:unbiasedconsistent} and
	$\Omega\succeq0$, we obtain $(n\widehat C_n)^+\to\Omega^+=\Omega$ almost surely.
\end{proof}
The PSD correction thus retains first-order consistency, but it generally does not preserve exact unbiasedness. Indeed,
$\widehat C_n^+-\widehat C_n$ is PSD and is nonzero whenever
$\widehat C_n$ has a negative eigenvalue. Hence
$\E(\widehat C_n^+)-\Cov(\widehat{\boldsymbol T}_n)$ is PSD, and it is
nonzero whenever $\widehat C_n$ is indefinite with positive probability.
When the estimate must be inverted, a positive-definite (PD) correction
is needed as well.

\begin{remark}[PSD and positive-definite correction]\label{rem:pd}
	Write $C_n$ for the true finite-sample covariance matrix of the moment vector, for example $C_n=\Cov(\widehat{\boldsymbol T}_n)$. The eigenvalue truncation below is the nearest symmetric PSD matrix in Frobenius norm \citep{Higham1988}; related eigenvalue-constrained covariance estimation is discussed by \citet{LiuWangZhao2014}. Write $\mathcal S_n=n\widehat C_n=V_n\operatorname{diag}(d_{1n},\ldots,d_{pn})V_n^\T$. Its PSD correction replaces $d_{jn}$ by $d_{jn}^+=\max(d_{jn},0)$; since $C_n$ is PSD, projection also gives $\|\widehat C_n^+-C_n\|_F\le\|\widehat C_n-C_n\|_F$. For inversion, choose $\varepsilon_n>0$ with $\varepsilon_n\to0$ and define
	\begin{equation}\label{eq:pdregularize}
	\widehat C_n^{\mathrm{PD}}=\frac1n V_n\operatorname{diag}\{\max(d_{1n},\varepsilon_n),\ldots,\max(d_{pn},\varepsilon_n)\}V_n^\T.
	\end{equation}
	Then $\widehat C_n^{\mathrm{PD}}\succ0$ for any observed data set and $n\widehat C_n^{\mathrm{PD}}\to\Omega$ almost surely. If $\Omega\succ0$, the inverse of the corrected \emph{scaled} covariance converges to $\Omega^{-1}$. When $\Omega$ is singular, PD regularization does not make a consistent estimator of a nonexistent $\Omega^{-1}$. Neither PSD nor PD correction generally retains finite-sample unbiasedness.
	
The statistical properties of the two corrections can be summarized as follows.
\begin{enumerate}[label=(\alph*)]
	\item \emph{Consistency.} Both corrections preserve first-order consistency: $n\widehat C_n^+\to\Omega$ by Corollary~\ref{cor:psdcorr}, and $\|n\widehat C_n^{\mathrm{PD}}-n\widehat C_n^+\|_F\le p^{1/2}\varepsilon_n\to0$ because the two matrices differ only in eigenvalues below $\varepsilon_n$.
	\item \emph{Unbiasedness.} Neither correction is unbiased. Since $\widehat C_n^{\mathrm{PD}}\succeq\widehat C_n^+\succeq\widehat C_n$ in the Loewner order, both corrections shift the estimator upward, and the shift is nonzero on the event that $\widehat C_n$ is indefinite. Table~\ref{tab:estimators} shows that this bias is small for the PSD projection.
	\item \emph{Singular limit.} If $\Omega$ is singular, for example when some linear combination of the moments has a degenerate first-order projection, the corrected matrices remain consistent for $\Omega$. Their smallest eigenvalues, however, tend to zero, and inverse-covariance weighting is ill-posed in the limit. A generalized inverse or a reduced moment vector is then more appropriate than inversion of $\widehat C_n^{\mathrm{PD}}$.
	\item \emph{Role of $\varepsilon_n$.} The floor $\varepsilon_n$ guarantees invertibility for any observed data set and bounds the condition number of $n\widehat C_n^{\mathrm{PD}}$ by $\max_jd_{jn}/\varepsilon_n$; the condition $\varepsilon_n\to0$ ensures that the regularization vanishes asymptotically. A larger floor stabilizes the inverse at the cost of more upward bias. Our implementation uses $\varepsilon_n=n^{-1}p^{-1}\sum_j|d_{jn}|$, which is $O(n^{-1})$; in the Gaussian design of Section~\ref{subsec:reference}, its median is $0.020$ and $0.008$ for $n=20$ and $50$, about $1\%$ of the largest eigenvalue of $n\widehat C_n$.
\end{enumerate}
\end{remark}

\section{Numerical experiments}\label{sec:numerical}

We assess the computational efficiency of the quadratic-time algorithm
(Sections~\ref{subsec:comp} and~\ref{subsec:scaling}) and the finite-sample
behavior of the exactly unbiased covariance estimator
(Sections~\ref{subsec:finite} and~\ref{subsec:reference}). The
accuracy of the first-order ratio approximation is examined in
Appendix~\ref{subsec:ratiosim}. Throughout this section,
$\widehat{\boldsymbol M}_n=(\widehat{\boldsymbol T}_n^\T,\widehat\lambda_2(X),\widehat\lambda_2(Y))^\T$
denotes the eight-dimensional vector of the sample means, the four
sample L-comoments, and the two marginal sample L-scales,
and $C_n=\Cov(\widehat{\boldsymbol M}_n)$ denotes its exact
finite-sample covariance matrix. 

Sections~\ref{subsec:comp}
and~\ref{subsec:finite} use independent bivariate Gaussian pairs with
unit marginal variances and correlation $\rho=0.6$. These experiments
were run in R 4.5.2 on an Intel Core i5-5200U (2.20 GHz, two physical
cores/four logical processors), with 8 GB RAM and Windows 10 Education;
times are elapsed wall-clock times. For the timing comparison, both
methods use the same simulated dataset at each $n$; the Monte Carlo
study uses $R=1000$ independent samples per sample size. 

The
experiments in Section~\ref{subsec:reference} and Appendix~\ref{subsec:ratiosim}
were run with an independent vectorized implementation in Python~3.13
with NumPy~2.5 on a Linux server (Intel Xeon, 2.1\,GHz, two cores,
8\,GB RAM). This implementation reproduces direct enumeration of the
disjoint-subset estimator to within $4\times10^{-16}$ for $6\le n\le9$.

\subsection{Computational efficiency}\label{subsec:comp}
The direct reference method evaluates the disjoint-subset expression in Theorem~\ref{thm:unbiased}; the fast method evaluates the algebraically equivalent rank-prefix implementation in Proposition~\ref{prop:fast}. Both produce the full $8\times8$ estimate of $C_n$ for the vector $\widehat{\boldsymbol M}_n$. Table~\ref{tab:computation} reports elapsed times on the same computer.
\begin{table}[htbp]
	\centering
	\caption{Computation of the full $8\times8$ exactly unbiased covariance matrix.}
	\label{tab:computation}
	\begin{tabular}{rrrrr}
		\toprule
		$n$ & Direct (s) & Fast (s) & Speedup & Max. absolute difference \\
		\midrule
		20 & 0.2100 & 0.0367 & 5.73 & $8.98\times10^{-17}$ \\
		30 & 1.6400 & 0.0900 & 18.22 & $2.39\times10^{-16}$ \\
		50 & 22.5300 & 0.2600 & 86.65 & $2.68\times10^{-16}$ \\
		\bottomrule
	\end{tabular}
	\par\smallskip
	\begin{minipage}{0.96\linewidth}\footnotesize
		\emph{Note.} Each time is the median of three measurements, except for the direct method at $n=50$, measured once. The fast method was batched to reduce timer-resolution effects. Speedup is Direct/Fast. The final column is the largest entrywise absolute difference between the computed matrices; seed $20261017$ was used for the Gaussian samples.
	\end{minipage}
\end{table}

The observed speedup increases from $5.73$ at $n=20$ to $86.65$ at $n=50$, while numerical discrepancies stay below $3\times10^{-16}$. Thus, on these datasets, the fast algorithm returns the same covariance estimates to floating-point precision at substantially lower computational cost. The growing advantage is consistent with the $O(n^2)$ arithmetic complexity established in Proposition~\ref{prop:fast}; the three measured sizes alone do not constitute a complexity proof.

\subsection{Finite-sample behavior of the unbiased estimator}\label{subsec:finite}
For each $n\in\{20,30,50\}$, we computed in each of $R=1000$ replications both the unbiased covariance estimate $\widehat C_n^{(b)}$ and the corresponding moment vector $\widehat{\boldsymbol M}_n^{(b)}$. The mean of the covariance estimates was compared with the sample covariance of the replicated moment vectors,
\[
\overline C_n=R^{-1}\sum_{b=1}^R\widehat C_n^{(b)},\qquad
C_{n,\mathrm{MC}}=\frac{1}{R-1}\sum_{b=1}^R
(\widehat{\boldsymbol M}_n^{(b)}-\overline{\boldsymbol M}_n)(\widehat{\boldsymbol M}_n^{(b)}-\overline{\boldsymbol M}_n)^\T.
\]
Both quantities estimate the same finite-sample sampling covariance, but their difference is subject to Monte Carlo variation. Table~\ref{tab:mc_covariance} reports the root mean squared entrywise difference, the relative Frobenius difference $\|\overline C_n-C_{n,\mathrm{MC}}\|_F/\|C_{n,\mathrm{MC}}\|_F$, and the largest absolute standardized difference among the 36 distinct entries. For each entry, the standardized difference is $z=(\bar c-c)/\mathrm{se}$, where $\bar c$ is the entry of $\overline C_n$, $c$ is the corresponding entry of $C_{n,\mathrm{MC}}$, and $\mathrm{se}$ is the Monte Carlo standard error of $\bar c-c$. We also report the fraction of covariance estimates having a negative minimum eigenvalue.

\begin{table}[htbp]
	\centering
	\caption{Monte Carlo diagnostics for the exactly unbiased covariance estimator ($R=1000$, $\rho=0.6$).}
	\label{tab:mc_covariance}
	\begin{tabular}{rrrrrr}
		\toprule
		$n$ & Max. $|z|$ & RMS diff. & Rel. Frobenius diff. & Non-PSD (\%) & Median min.\ eigenvalue \\
		\midrule
		20 & 1.708 & 0.000926 & 0.0770 & 99.0 & $-5.957\times10^{-4}$ \\
		30 & 2.571 & 0.000464 & 0.0611 & 82.2 & $-6.909\times10^{-5}$ \\
		50 & 2.885 & 0.000400 & 0.0920 & 29.3 & $\phantom{-}1.454\times10^{-5}$ \\
		\bottomrule
	\end{tabular}
	\par\smallskip
	\begin{minipage}{0.96\linewidth}\footnotesize
		\emph{Note.} Max. $|z|$ is the largest standardized Monte Carlo difference among the 36 distinct covariance entries; it is descriptive rather than a simultaneous test. Non-PSD is the observed percentage of indefinite estimates (negative minimum eigenvalue). The simulation seed was $20261018$.
	\end{minipage}
\end{table}

The standardized discrepancies are moderate (maximum absolute values $1.71$, $2.57$, and $2.88$), and the simulation findings do not contradict the exact unbiasedness established in Theorem~\ref{thm:unbiased}. The RMS differences decrease over the three sample sizes, but the relative Frobenius differences do not decrease monotonically; the simulation should not be interpreted as demonstrating a general rate of convergence. In particular, the empirical reference covariance is itself estimated with Monte Carlo error.

The proportion of non-PSD unbiased estimates drops from $99.0\%$ at $n=20$ to $29.3\%$ at $n=50$ in this Gaussian design. The median minimum eigenvalue changes from $-5.96\times10^{-4}$ to $1.45\times10^{-5}$. This illustrates the distinction between finite-sample unbiasedness and the samplewise PSD property, and motivates the eigenvalue corrections in Remark~\ref{rem:pd}. Such corrections do not, in general, preserve unbiasedness. These empirical frequencies are specific to the distribution, estimator, and sample sizes considered.

\subsection{Comparison with high-precision reference covariances}\label{subsec:reference}
In Table~\ref{tab:mc_covariance}, $\overline C_n$ is compared with an empirical covariance computed from the same $R=1000$ replications, so the reference itself carries Monte Carlo error of the same order as the quantity being assessed. We therefore ran a second study in which the reference covariance is computed much more precisely and in which three covariance estimators are compared. Three designs were considered:
\begin{enumerate}[label=(\roman*)]
	\item independent variables, $X\sim\mathrm{Exp}(1)$ and $Y\sim\mathrm{Gumbel}(0,1)$, for which Theorem~\ref{thm:indep} gives the exact $4\times4$ L-comoment block;
	\item the Gaussian design above with $\rho=0.6$; and
	\item a Gaussian copula with correlation $0.6$ and generalized extreme-value margins with location $0$, scale $1$, and shape $\xi=0.2$, which have a heavy upper tail and finite variance.
\end{enumerate}
Design~(iii) represents the skewed, heavy-tailed data that motivate L-moment methods. For each design and $n\in\{20,30,50\}$, the reference covariance $C_n$ was computed from $10^6$ independent replications of $\widehat{\boldsymbol M}_n$; this requires only the sample moments, not their covariance estimators. In design~(i), the reference agreed with Theorem~\ref{thm:indep} to within $0.3\%$ of the corresponding standard deviations for all ten distinct entries of the $4\times4$ block, which also confirms the closed form numerically.

We then generated $R=20000$ further samples for each setting and computed three estimators: the unbiased estimator $\widehat C_n$, its PSD projection $\widehat C_n^+$, and the projection-based estimator $\widehat\Omega_n/n$ of Theorem~\ref{thm:projectionest}. For a generic estimator $\widetilde C_n$, Table~\ref{tab:estimators} reports the largest absolute standardized difference between the Monte Carlo mean of $\widetilde C_n$ and $C_n$ over the 36 distinct entries, with standard errors that combine the Monte Carlo errors of both quantities; the relative bias (RB); and the relative root mean squared error (RMSE). To make entries of different scales comparable, let $\Psi_n=\operatorname{diag}(C_n)$ and, for a $p\times p$ matrix $A$ with $p=8$, let $\|A\|_*=\|\Psi_n^{-1/2}A\Psi_n^{-1/2}\|_F$ be the Frobenius norm after each entry has been divided by the product of the corresponding reference standard deviations. Then
\begin{equation*}
\mathrm{RB}=100\times\frac{\|\E(\widetilde C_n)-C_n\|_*}{\|C_n\|_*},\qquad
\mathrm{RMSE}=\frac{\{\E(\|\widetilde C_n-C_n\|_*^2)\}^{1/2}}{\|C_n\|_*},
\end{equation*}
where the expectations are Monte Carlo averages over the $R$ samples. Thus RB measures the systematic error of an estimator, and RMSE measures its total error, both relative to the size of $C_n$; an RMSE of one means that the typical estimation error is as large as $C_n$ itself on the standardized scale.

\begin{table}[htbp]
	\centering
	\caption{Comparison of covariance estimators with high-precision reference covariances ($R=20000$; reference from $10^6$ replications). RB and RMSE are the relative bias (\%) and the relative root mean squared error defined in Section~\ref{subsec:reference}.}
	\label{tab:estimators}
	\footnotesize\setlength{\tabcolsep}{4pt}
	\begin{tabular}{lr c ccc cc ccc}
		\toprule
		&&& \multicolumn{3}{c}{Unbiased $\widehat C_n$} & \multicolumn{2}{c}{PSD $\widehat C_n^+$} & \multicolumn{3}{c}{Projection $\widehat\Omega_n/n$}\\
		\cmidrule(lr){4-6}\cmidrule(lr){7-8}\cmidrule(lr){9-11}
		Design & $n$ & Non-PSD (\%) & Max.\ $|z|$ & RB & RMSE & Max.\ $|z|$ & RB & Max.\ $|z|$ & RB & RMSE\\
		\midrule
		Independent & 20 & 99.4 & 2.32 & 1.15 & 1.399 & 5.08 & 1.71 & 29.44 & 8.54 & 1.194 \\
		Independent & 30 & 81.9 & 1.30 & 0.48 & 1.127 & 1.93 & 0.66 & 20.44 & 5.28 & 1.017 \\
		Independent & 50 & 18.4 & 2.12 & 0.70 & 0.853 & 2.12 & 0.70 & 14.67 & 3.37 & 0.803 \\
		\addlinespace
		Gaussian & 20 & 98.9 & 2.62 & 0.69 & 0.929 & 6.07 & 1.37 & 25.15 & 7.23 & 0.784 \\
		Gaussian & 30 & 81.4 & 2.18 & 0.57 & 0.728 & 2.23 & 0.58 & 18.96 & 4.66 & 0.653 \\
		Gaussian & 50 & 29.7 & 3.07 & 0.56 & 0.545 & 3.07 & 0.56 & 13.37 & 2.99 & 0.512 \\
		\addlinespace
		GEV copula & 20 & 99.4 & 1.42 & 1.81 & 2.898 & 1.61 & 1.80 & 10.68 & 12.92 & 2.432 \\
		GEV copula & 30 & 86.4 & 1.14 & 0.89 & 2.636 & 1.14 & 1.00 & 6.67 & 7.04 & 2.345 \\
		GEV copula & 50 & 32.3 & 0.69 & 0.49 & 2.188 & 0.68 & 0.50 & 5.29 & 4.55 & 2.042 \\
		\bottomrule
	\end{tabular}
	\par\smallskip
	\begin{minipage}{0.96\linewidth}\footnotesize
		\emph{Note.} Max.\ $|z|$ is taken over the 36 distinct entries of the $8\times8$ matrix. In the independent design, the largest absolute standardized difference between the mean of $\widehat C_n$ and the exact block of Theorem~\ref{thm:indep} over its ten distinct entries is $2.05$, $0.94$, and $2.67$ for $n=20,30,50$. Seed $20261018$ (NumPy generator).
	\end{minipage}
\end{table}

The unbiased estimator performs as the theory predicts. It shows no detectable bias: its largest standardized difference is at most $3.07$ over 36 correlated entries, and its relative bias, which is itself subject to Monte Carlo error, is below $2\%$ in every setting, including the heavy-tailed design. In the independent design, its mean also agrees with the exact block of Theorem~\ref{thm:indep}. The frequencies of non-PSD estimates in the Gaussian design ($98.9\%$, $81.4\%$, and $29.7\%$) agree closely with those in Table~\ref{tab:mc_covariance}, which were obtained with the \textsf{R} implementation and different random numbers. Projection onto the PSD cone removes the indefiniteness at the cost of a small bias at $n=20$, with standardized differences up to about $6$ and relative bias below $2\%$, and it changes the estimator very little for $n\ge30$.

The projection-based estimator $\widehat\Omega_n/n$, in contrast, is systematically biased, because it targets only the leading term $\Omega/n$ of \eqref{eq:augcov}. At $n=20$ its diagonal entries understate the true variances by up to about $14\%$; its relative bias decreases from $7$--$13\%$ at $n=20$ to $3$--$5\%$ at $n=50$, and its largest standardized differences remain between $5$ and $29$. Nevertheless, it has the smaller relative RMSE in all nine settings, by $6$--$16\%$ relative to $\widehat C_n$, because it is less variable. Table~\ref{tab:estimators} therefore does not show that the unbiased estimator is uniformly better in small samples; the two estimators exhibit a bias--variance trade-off.

If the aim is to minimize the bias of the estimated finite-sample covariance, the unbiased estimator, or its PSD projection, is preferable; this is the relevant criterion when standard errors should not be understated or when covariance estimates are averaged over many sites or samples, because random errors then average out whereas the bias of $\widehat\Omega_n/n$ does not. If the criterion is the squared-error loss of the whole covariance matrix for a single sample, the projection-based estimator can be preferable, as it was in all nine settings considered here.

The relative RMSE values are large for all three estimators, especially in the GEV-copula design, where they range from $2.0$ to $2.9$; values above one also occur in the independent design for $n\le30$. This reflects the intrinsic difficulty of estimating a covariance matrix of L-comoments from one small sample: each entry of $\widetilde C_n$ is a second moment of the moment vector, so its sampling variability depends on fourth moments of the data. For GEV margins with shape $\xi$, the fourth moment is finite only when $\xi<1/4$, so $\xi=0.2$ lies close to this boundary, and the estimators have heavy-tailed sampling distributions whose occasional large values dominate the mean squared error. The RMSE decreases with $n$ in every design, and the ranking of the estimators is the same as under lighter tails. In heavy-tailed applications, covariance estimates from samples of this size should therefore be interpreted with caution.

\subsection{Scaling of the fast algorithm}\label{subsec:scaling}
Table~\ref{tab:computation} is limited to $n\le50$ because the direct method becomes impractical for larger samples. To examine the growth of the fast algorithm alone, Table~\ref{tab:scaling} reports elapsed times of a vectorized NumPy implementation of Proposition~\ref{prop:fast} for $n$ up to $4000$; each time is the median of five measurements ($n\le1000$) or three measurements ($n\ge2000$).

\begin{table}[htbp]
	\centering
	\caption{Elapsed time of the fast algorithm for the full $8\times8$ unbiased covariance matrix (vectorized NumPy implementation).}
	\label{tab:scaling}
	\begin{tabular}{rrrrrrr}
		\toprule
		$n$ & 100 & 200 & 500 & 1000 & 2000 & 4000\\
		\midrule
		Time (s) & 0.010 & 0.047 & 0.336 & 1.543 & 8.489 & 44.02\\
		Prefix tables (MB) & 0.5 & 1.8 & 11.5 & 45.9 & 183.3 & 732.8\\
		\bottomrule
	\end{tabular}
\end{table}

The least-squares slope of log time on log $n$ is $2.18$ for $100\le n\le1000$ and $2.26$ over all sizes, close to the quadratic arithmetic complexity established in Proposition~\ref{prop:fast}. The slight excess at the largest sizes reflects memory traffic in the large prefix tables. The six prefix tables require $48(n+1)^2$ bytes, about $46$\,MB at $n=1000$ and $0.73$\,GB at $n=4000$. The vectorized implementation also stores pair-level quantities, and its peak memory use was about $4.9$\,GB at $n=4000$; a loop over pairs, as in the \textsf{R} implementation, avoids this pair-level storage. The algorithm is therefore practical well beyond the record lengths typical of L-moment applications, whereas direct enumeration is already slow at $n=50$.

\section{Discussion}\label{sec:discussion}
The results of this study are potentially relevant to multivariate regional frequency
analysis \citep{ChebanaOuarda2007,AkinsojiEtAl2026}. In the discordancy and homogeneity tests of
\citet{ChebanaOuarda2007}, the dispersion of at-site L-comoment
coefficients is weighted by record length, and its null distribution
is obtained by simulating homogeneous regions from a fitted
bivariate model, whose choice the authors note is subjective. The
present estimators give the sampling covariance of each site's
L-comoments from the site's own data, without a parametric model. They
could potentially provide precision weights in place of weights based
on record length, and they may serve as a basis for a simulation-free
approximation to the null distribution of such statistics. These tests also use fourth-order
L-comoment coefficients; the projection estimator covers them
(Corollary~\ref{cor:order4proj}), whereas the exactly unbiased
estimator of order four requires the algorithmic extension noted in
Remark~\ref{rem:order4}. These order-four results follow directly from
fixed-order U-statistic theory and the leave-one-out identity, and we
include them as a supplementary extension because of this use in practice. Spatial dependence between sites, studied by
\citet{Simkova2017}, is outside our assumption of independent pairs.

When $X$ and $Y$ are measured on the same scale, the symmetric
component $S_m=(\lambda_m^{[XY]}+\lambda_m^{[YX]})/2$ and the
difference between the two directions $D_m=(\lambda_m^{[XY]}-\lambda_m^{[YX]})/2$ may
also be of interest; symmetric Gini constructions are studied by
\citet{SangDangSang2016}. Following \citet{SerflingXiao2007}, we
treated the ordered pair $(\lambda_m^{[XY]},\lambda_m^{[YX]})$ as the
basic object. The vector $(\widehat S_2,\widehat D_2,\widehat S_3,\widehat D_3)^\T$
equals $\mathsf A\widehat\bL$, where $\mathsf A$ is block diagonal with
two blocks $\frac12\bigl(\begin{smallmatrix}1&1\\1&-1\end{smallmatrix}\bigr)$.
Its exact covariance is therefore $\mathsf A\Cov(\widehat\bL)\mathsf A^\T$
by Theorem~\ref{thm:exact}; as noted in Section~\ref{subsec:indep}, ignoring
the correlation between the two directions overstates the variance of
$\widehat D_m$. Under exchangeability, $(X,Y)\overset d=(Y,X)$, we have
$\E(\widehat D_m)=0$ and $\Cov(\widehat S_m,\widehat D_s)=0$ for
$m,s\in\{2,3\}$, which could serve as the basis of a test of
exchangeability. Related tests based on the difference between the
two directional Gini coefficients, such as the test for time
reversibility of \citet{ShelefSchechtman2019}, have relied on bootstrap
approximations; for independent pairs, the exact covariance offers an
alternative.

For routine asymptotic standard errors, the projection-based PSD estimator
is a natural default. When finite-sample bias should be avoided, the
unbiased estimator or its PSD projection is preferable, whereas the
projection-based estimator can be preferable under squared-error loss for
the whole covariance matrix (Section~\ref{subsec:reference}). For the ratio $\widehat\tau_3$, the first-order
approximation can overstate the variance under heavy-tailed margins, by
up to $29\%$ at $n=20$ in our GEV design (Appendix~\ref{subsec:ratiosim}). The exact
covariance results extend those of \citet{ElamirSeheult2004} for
univariate L-moments to L-comoments, and the rank-prefix
algorithm of Section~\ref{sec:fast} makes the corresponding unbiased
estimator practical for large samples. PSD and PD regularizations based
on eigenvalue correction \citep{Higham1988,LiuWangZhao2014} are available
when a covariance-matrix form or an inverse is required, although they
generally sacrifice exact unbiasedness. Downstream inference based on
multivariate L-moment estimation also requires identification and
appropriate moment Jacobians. Inverse covariance weighting
\citep{ShinShinPark2026SERRA} further requires a nonsingular limiting
covariance.

The results assume independent observation pairs and continuous
margins, with unrestricted dependence within each pair. Higher-order
L-comoments, L-comoment matrices of more than two variables, and trimmed
L-comoments \citep{ElamirSeheult2003,Hosking2007} are again fixed-order
U-statistics. The overlap formula, exact unbiasedness, and consistency
therefore extend directly; only the conditional kernels and the
rank-prefix tables must be redeveloped. Ties can be handled with a
kernel convention, such as $\sgn(0)=0$ and averaging over tied
orderings. This convention keeps the above properties but changes the
targets, the closed forms, and the algorithm. For the $r$ largest observations
of blocks \citep{Tawn1988,BusababodhinShinPark2025,BusababodhinShinPark2026},
the present results apply when the blocks are independent, because the
dependence among the order statistics within a block is allowed.

\section{Conclusion}\label{sec:conclusion}
We derived the exact finite-sample covariance structure of sample
L-comoments in both directions, for bivariate
samples with unrestricted dependence within each pair. Explicit
conditional kernels give every entry of the exact $4\times4$ sampling
covariance matrix and its $6\times6$ extension that includes the sample
means. They also yield a canonical Hoeffding decomposition of this
covariance. When $X$ and $Y$ are independent, the covariance matrix is fully
explicit and shows that the estimators of the two directions are
correlated even when both population L-comoments vanish. We considered
two estimators of this sampling covariance matrix. The first is exactly
unbiased for the finite-sample covariance, is first-order consistent,
and can be corrected to be PSD or PD. The second is a projection
estimator of the leading asymptotic covariance, which is consistent and
PSD for any observed data set.

An overlap inclusion--exclusion identity combined with rank-prefix sums
evaluates the unbiased estimator in $O(n^2)$ time instead of $O(n^6)$.
The numerical experiments confirmed the exact formulas and the agreement
of the fast and direct algorithms. For samples of size $20$--$50$, the
unbiased estimator had no detectable bias, whereas the projection
estimator understated the covariance by several percent. The projection
estimator nevertheless had the smaller root mean squared error, so the
choice between the two estimators is a bias--variance trade-off. The
first-order approximation for L-comoment ratios was accurate except for
$\widehat\tau_3$ under heavy-tailed margins, where it overstated the
variance. These results may be useful in subsequent studies of
inference based on L-comoments, for example in tests of independence or
exchangeability, in the discordancy and homogeneity tests of regional
frequency analysis, and in multivariate L-moment estimation with
inverse-covariance weighting.

{\small\linespread{1.0}\selectfont
\section*{Declarations}
\noindent\textbf{Competing interests.} The authors declare that they have no known competing financial interests or personal relationships that could have appeared to influence the work reported in this paper.\\[4pt]
\noindent\textbf{Funding.} 
The research of P.S. was supported by Khon Kaen University, Thailand.\\[4pt]
\noindent\textbf{Author contributions.} Palakorn Seenoi: conceptualization, formal analysis, software, numerical experiments, and drafting. Jeong-Soo Park: conceptualization, methodology, supervision, mathematical verification, and writing--review and editing.\\[4pt]
\noindent\textbf{Data availability.} No external observational data were used. All numerical results are based on simulated bivariate observations described in the paper.\\[4pt]
\noindent\textbf{Code availability.} \textsf{R} code for the numerical experiments and covariance computation is available at \url{https://github.com/palakorn-seenoi/sample_Lcomom}. \\[4pt]
\noindent\textbf{Declaration of generative AI use.} ChatGPT (OpenAI) and Claude (Anthropic) were used for language editing, technical discussion, and mathematical derivations. The authors are responsible for reviewing and verifying all derivations, results, citations, and the final manuscript. \\[4pt]
\noindent\textbf{ORCID.} Palakorn Seenoi: \href{https://orcid.org/0000-0001-9374-070X}{0000-0001-9374-070X}; Jeong-Soo Park: \href{https://orcid.org/0000-0002-8460-4869}{0000-0002-8460-4869}.
\par}

\appendix
\section*{Appendix}
\section{Preliminaries on U-statistics and L-comoments}\label{app:ustat}
We collect standard facts used repeatedly in the paper, without proofs.

\paragraph{Results of \citet{SerflingXiao2007}.} For $(X,Y)$ with
finite mean and continuous margins, the $k$th L-comoment of $X$ with
respect to $Y$ is
$\lambda_k^{[XY]}=\Cov[X,P_{k-1}\{F_Y(Y)\}]$, $k\ge2$, which equals
\eqref{eq:poplambda} because $\E[P_{k-1}\{F_Y(Y)\}]=0$. The following
facts are used in this paper.
\begin{enumerate}[label=(\roman*)]
\item \emph{Invariance} (their equation~(11)). $\lambda_k^{[XY]}$ is
translation invariant and scale equivariant in $X$, and translation and
scale invariant in $Y$.
\item \emph{Special cases.} If $Y=X$, $\lambda_k^{[XY]}$ is the
univariate L-moment $\lambda_k(X)$. If $X$ and $Y$ are independent,
$\lambda_k^{[XY]}=0$ for all $k\ge2$.
\item \emph{Concomitant representation} (their Proposition~6).
\begin{equation*}
\lambda_k^{[XY]}=k^{-1}\sum_{j=0}^{k-1}(-1)^j\binom{k-1}j
\E\bigl(X_{[k-j:k]}\bigr);
\end{equation*}
the cases $k=2,3,4$ are
\eqref{eq:popconcomitant}.
\item \emph{Unbiased estimation} (their Proposition~7 and
equation~(26)). The sample L-comoment, a weighted sum of the sample
concomitants with the weights of univariate sample L-moments, is an
unbiased estimator of $\lambda_k^{[XY]}$ and is also a U-statistic;
for $k=2,3$ this is \eqref{eq:rankstat}.
\item \emph{Asymptotic normality} (their Proposition~8). Under finite
second moments of $X$, the vector of sample L-comoments of orders
$2,\ldots,k$ is asymptotically normal, with covariance matrix of order
$n^{-1}$ given by the first-projection covariances; this agrees with
the leading term $B_1/n$ in Theorem~\ref{thm:Bdecomp}.
\end{enumerate}
Our contribution is the exact finite-sample covariance of these
estimators, its estimation, and its efficient computation.

\paragraph{U-statistics.} See \citet{Hoeffding1948},
\citet[Chapter~5]{Serfling1980}, \citet{Lee1990}, and
\citet{vanDerVaart1998}. Let $Z_1,\ldots,Z_n$ be iid, and let $h$ be a
symmetric kernel of order $m$ with a finite second moment. Write
$\theta=\E\{h(Z_1,\ldots,Z_m)\}$. The complete U-statistic
$U_n=\binom nm^{-1}\sum_{|I|=m}h(Z_I)$ is an unbiased estimator of
$\theta$; a sample mean is the case $m=1$. For $1\le c\le m$, put
\begin{equation*}
h_c(z_1,\ldots,z_c)=\E\{h(Z_1,\ldots,Z_m)\mid Z_1=z_1,\ldots,Z_c=z_c\},
\end{equation*}
so that $h_m=h$ and $\E\{h_c(Z_1,\ldots,Z_c)\}=\theta$.

\paragraph{Overlap formula.} Let $U_n$ and $V_n$ be U-statistics with
kernels $h$ of order $m$ and $g$ of order $s$. Then
\begin{equation}\label{eq:appoverlap}
\Cov(U_n,V_n)=\sum_{c=1}^{\min(m,s)}
\frac{\binom mc\binom{n-m}{s-c}}{\binom ns}\,
\Cov\{h_c(Z_1,\ldots,Z_c),g_c(Z_1,\ldots,Z_c)\}.
\end{equation}
Terms with disjoint index sets contribute zero, and a term whose index
sets share $c$ observations contributes the covariance of the
conditional kernels given those observations \citep{Hoeffding1948}.

\paragraph{Hoeffding decomposition.} Define the canonical kernels
recursively by $h^{(1)}(z_1)=h_1(z_1)-\theta$ and, for $c\ge2$,
\begin{equation*}
h^{(c)}(z_1,\ldots,z_c)=h_c(z_1,\ldots,z_c)-\theta
-\sum_{j=1}^{c-1}\sum_{1\le i_1<\cdots<i_j\le c}h^{(j)}(z_{i_1},\ldots,z_{i_j}).
\end{equation*}
Each $h^{(c)}$ has zero conditional mean given any proper subset of its
arguments. With $U_n^{(c)}=\binom nc^{-1}\sum_{|I|=c}h^{(c)}(Z_I)$,
\begin{equation}\label{eq:apphoeffding}
U_n-\theta=\sum_{c=1}^m\binom mc U_n^{(c)},
\end{equation}
and the terms are uncorrelated across $c$. In particular,
$\Var(U_n)=\sum_{c=1}^m\binom mc^2\binom nc^{-1}\Var\{h^{(c)}\}$, whose
leading term is $m^2\Var\{h_1(Z_1)\}/n$ \citep{Lee1990}.

\paragraph{Limit theorems.} By the strong law of large numbers for
U-statistics, $U_n\to\theta$ almost surely when $\E|h|<\infty$
\citep[Chapter~5]{Serfling1980}. By the central limit theorem,
$\sqrt n(U_n-\theta)\xrightarrow{d}N(0,m^2\Var\{h_1(Z_1)\})$; the
multivariate version holds jointly for finitely many U-statistics, with
asymptotic covariance $ms\Cov\{h_1(Z_1),g_1(Z_1)\}$ between kernels of
orders $m$ and $s$ \citep[Chapter~5]{Serfling1980}. These results,
combined with the delta method \citep{vanDerVaart1998}, give the limits
of smooth functions such as ratios.

\paragraph{Leave-one-out identity.} Let
$\widehat h_1(Z_i)=\binom{n-1}{m-1}^{-1}\sum_{I\ni i,\,|I|=m}h(Z_I)$ be
the average of the kernel over the subsets containing observation $i$,
and let $U_{n-1}^{(-i)}$ be the U-statistic computed from the sample
without observation $i$. Then
\begin{equation}\label{eq:apploo}
\widehat h_1(Z_i)=\frac{\binom nm U_n-\binom{n-1}m U_{n-1}^{(-i)}}{\binom{n-1}{m-1}},
\qquad i=1,\ldots,n,
\end{equation}
because the subsets containing $i$ are those not contained in the
sample without $i$.

\paragraph{Disjoint-subset principle.} If $I$ and $J$ are disjoint,
$h(Z_I)$ and $g(Z_J)$ are independent, so
$\E\{h(Z_I)g(Z_J)\}=\E(h)\E(g)$. Averaging such products over all
disjoint pairs $(I,J)$ therefore gives an unbiased estimator of the
product of means, and subtracting it from $U_nV_n$ gives an unbiased
estimator of $\Cov(U_n,V_n)$ \citep{Hoeffding1948}.

\paragraph{Inclusion--exclusion.} For finite index sets $I$ and $J$,
$\mathbf 1\{I\cap J=\varnothing\}=\sum_{K\subseteq I\cap J}(-1)^{|K|}$.
This identity converts sums over disjoint subsets into sums over
overlapping ones, as used in Proposition~\ref{prop:fast}.

\section{Concomitant coefficients, conditional kernels, and calculations under independence}\label{app:general}

\subsection{Concomitant representation and means of the kernels}\label{app:kernelmean}
Put $U=F_Y(Y)$. Since $Y$ is continuous, the pair $(X_j,Y_j)$ has the
$r$th smallest $Y$ value among $k$ pairs with conditional probability
$\binom{k-1}{r-1}U_j^{r-1}(1-U_j)^{k-r}$ given $(X_j,Y_j)$. Hence
\begin{equation}\label{eq:concomitantmean}
\E\bigl(X_{[r:k]}\bigr)=k\binom{k-1}{r-1}\E\bigl\{XU^{r-1}(1-U)^{k-r}\bigr\}
\end{equation}
\citep[Section~3.2]{SerflingXiao2007}. Substituting
\eqref{eq:concomitantmean} into the right-hand sides of
\eqref{eq:popconcomitant} gives, inside $\E\{X\,\cdot\,\}$,
\begin{align*}
U-(1-U)&=2U-1=P_1(U),\\
U^2-4U(1-U)+(1-U)^2&=6U^2-6U+1=P_2(U),\\
U^3-9U^2(1-U)+9U(1-U)^2-(1-U)^3&=20U^3-30U^2+12U-1=P_3(U),
\end{align*}
for $k=2,3,4$, respectively. These are the definitions in
\eqref{eq:poplambda}, which proves \eqref{eq:popconcomitant}. The
kernels $H_2^{[XY]}$ and $H_3^{[XY]}$ in \eqref{eq:Htwo} and
\eqref{eq:Hthree} are the quantities inside the expectations in
\eqref{eq:popconcomitant}, evaluated at $m$ distinct pairs; therefore
$\E\{\bH_m(Z_I)\}=\boldsymbol\lambda_m$ for $m=2,3$. The $[YX]$
direction follows by exchanging the coordinates.

\subsection{Complete U-statistics and explicit rank coefficients}
Using the standard concomitant representation \citep{Yang1977,DavidNagaraja2003,SerflingXiao2007}, fix the order-two kernel in \eqref{eq:Htwo}, and sort all $n$ values of $Y$ increasingly. An observation with rank $q$ appears as the larger $Y$ coordinate in $q-1$ different pairs and as the smaller coordinate in $n-q$ different pairs. Its coefficient in the average of the order-two kernels therefore equals
\begin{equation*}
\frac1{{n\choose2}}\left\{\frac{q-1}{2}-\frac{n-q}{2}\right\}
=\frac{2q-n-1}{n(n-1)}=w_{2,q}.
\end{equation*}
For the order-three kernel \eqref{eq:Hthree}, an observation of rank $q$ can appear as the smallest, middle, or largest among a selected triple. The numbers of such triples are, respectively,
\begin{equation*}
{n-q\choose2},\qquad(q-1)(n-q),\qquad{q-1\choose2}.
\end{equation*}
Its contribution in the three positions is, respectively, $1/3,-2/3,1/3$ times its measured value. Hence the coefficient is
\begin{align*}
w_{3,q}&=\frac{{n-q\choose2}-2(q-1)(n-q)+{q-1\choose2}}{3{n\choose3}}
=\frac{6q^2-6(n+1)q+(n+1)(n+2)}{n(n-1)(n-2)}.
\end{align*}
This proves the concomitant representation \eqref{eq:rankstat}, including the exact coefficients. The sum of the coefficients is zero for each order because the sample kernel is unchanged by adding the same constant to all measured values.

\subsection{Elementary forms and a numerical example}\label{app:example}
In elementary terms, \eqref{eq:completeU} and \eqref{eq:rankstat}
amount to the following. For each pair of observations, take the $x$
value of the pair with the larger $y$ value minus the $x$ value of the
pair with the smaller $y$ value; $\widehat\lambda_2^{[XY]}$ is half the
average of these differences over all $\binom n2$ pairs. For each triple
of observations, ordered by their $y$ values, form
$x_{[1]}-2x_{[2]}+x_{[3]}$; $\widehat\lambda_3^{[XY]}$ is one third of
the average of these quantities over all $\binom n3$ triples.
Equivalently, in the probability-weighted-moment form of
\citet{Hosking1990}, put
\begin{equation}\label{eq:pwm}
b_0=\frac1n\sum_{q=1}^nX_{[q:n]},\quad
b_1=\frac1n\sum_{q=1}^n\frac{q-1}{n-1}X_{[q:n]},\quad
b_2=\frac1n\sum_{q=1}^n\frac{(q-1)(q-2)}{(n-1)(n-2)}X_{[q:n]}.
\end{equation}
Then $\widehat\lambda_2^{[XY]}=2b_1-b_0$ and
$\widehat\lambda_3^{[XY]}=6b_2-6b_1+b_0$, exactly as for the univariate
sample L-moments, except that the order statistics of $X$ are replaced
by the concomitants $X_{[q:n]}$ \citep{SerflingXiao2007,ChebanaOuarda2007}.

For example, let $n=5$ and $(x_i,y_i)=(3,1.3)$, $(8,0.5)$, $(5,2.6)$, $(12,3.4)$,
$(6,2.1)$. Sorting by $y$ gives the concomitants
$(X_{[1:5]},\ldots,X_{[5:5]})=(8,3,6,5,12)$. Then
$b_0=34/5=6.8$, $b_1=(0\cdot8+1\cdot3+2\cdot6+3\cdot5+4\cdot12)/20=3.9$,
and $b_2=(0\cdot8+0\cdot3+2\cdot6+6\cdot5+12\cdot12)/60=3.1$. Hence
$\widehat\lambda_2^{[XY]}=2(3.9)-6.8=1$ and
$\widehat\lambda_3^{[XY]}=6(3.1)-6(3.9)+6.8=2$. The weights in
\eqref{eq:rankstat} are $(w_{2,q})=(-4,-2,0,2,4)/20$ and
$(w_{3,q})=(12,-6,-12,-6,12)/60$, which give the same values, and so
does direct averaging of the kernels over the $10$ pairs and the $10$
triples.

\subsection{Conditional order-two kernel}
Fix $(x,y)$ and let $(X',Y')$ be an independent pair. Since $Y$ is continuous,
\begin{align*}
2M_{2,1}^{[XY]}(x,y)
&=\E((x-X')\sgn(y-Y'))\\
&=x\E(\ind(Y'<y)-\ind(Y'>y))
 -\E(X'\ind(Y'<y))+\E(X'\ind(Y'>y))\\
&=x\{2F_Y(y)-1\}+\mu_X-2Q_X(y).
\end{align*}
Dividing by two yields \eqref{eq:M21main}. This calculation also verifies that $\E(M_{2,1}^{[XY]}(Z))=\lambda_2^{[XY]}$.

\subsection{Conditional order-three kernel with two fixed pairs}
Fix $(x_\ell,y_\ell)$ and $(x_h,y_h)$, $y_\ell<y_h$, and let the third observation be $(X',Y')$. Define $u_\ell=F_Y(y_\ell)$ and $u_h=F_Y(y_h)$. The three possible relative positions of $Y'$ give
\begin{align*}
Y'<y_\ell&:\quad 3H_3=X'-2x_\ell+x_h,\\
y_\ell<Y'<y_h&:\quad 3H_3=x_\ell-2X'+x_h,\\
Y'>y_h&:\quad 3H_3=x_\ell-2x_h+X'.
\end{align*}
The conditional expectation of $3H_3$ can therefore be written as
\begin{align*}
3M_{3,2}^{[XY]}
={}&u_\ell(-2x_\ell+x_h)+(u_h-u_\ell)(x_\ell+x_h)+(1-u_h)(x_\ell-2x_h)\\
&+\E(X'\ind(Y'<y_\ell))-2\E(X'\ind(y_\ell<Y'<y_h))
 +\E(X'\ind(Y'>y_h)).
\end{align*}
Collect the $x_\ell$ coefficient:
$-2u_\ell+(u_h-u_\ell)+(1-u_h)=1-3u_\ell$.
The $x_h$ coefficient is
$u_\ell+(u_h-u_\ell)-2(1-u_h)=3u_h-2$.
For the terms containing $X'$, by continuity,
\begin{align*}
&Q_X(y_\ell)-2\{Q_X(y_h)-Q_X(y_\ell)\}+\{\mu_X-Q_X(y_h)\}\\
&\qquad=\mu_X-3\{Q_X(y_h)-Q_X(y_\ell)\}.
\end{align*}
Combining all three coefficients and dividing by three gives exactly \eqref{eq:M32main}.

\subsection{Conditional order-three kernel with one fixed pair}
We obtain the first projection by conditioning directly on the relative
ranks of the three observations, as in the preceding subsection. Fix
$Z_1=z=(x,y)$, write $u_y=F_Y(y)$, and for a generic pair $(X,Y)$ put
$W=F_Y(Y)$; for the pair $Z_2$, $W=F_Y(Y_2)$.
In $3H_3^{[XY]}(Z_1,Z_2,Z_3)$, the coefficient of $x$
is $1$, $-2$, or $1$ according as $y$ is smallest, middle,
or largest among the three $Y$ values. Since $Y_2,Y_3$ are iid,
its conditional mean equals
\begin{equation*}
(1-u_y)^2-4u_y(1-u_y)+u_y^2
=6u_y^2-6u_y+1=P_2(u_y).
\end{equation*}
Next fix $Z_2=(X_2,Y_2)$ and average over the position of $Y_3$.
For $Y_2<y$, the coefficient of $X_2$ in $3H_3^{[XY]}$
has conditional expectation $-2W+(1-W)=1-3W$;
for $Y_2>y$, it is $W-2(1-W)=3W-2$.
The same contribution arises from $X_3$ by symmetry. Consequently,
\begin{align*}
3M_{3,1}^{[XY]}(x,y)
={}&xP_2(u_y)
 +2\E\!\left[X\left\{(1-3W)\ind(Y<y)
                     +(3W-2)\ind(Y>y)\right\}\right]\\
={}&xP_2(u_y)+\E\bigl[X(2-6W)\bigr]
       +\E\bigl[X(12W-6)\ind(Y>y)\bigr].
\end{align*}
Here $R_X$ and $K_X$ are defined in \eqref{eq:RKmain}.
Since $P_2'(w)=12w-6$ and
$2\lambda_3^{[XY]}=\E\{X(12W^2-12W+2)\}$,
\begin{equation*}
\E\bigl[X(2-6W)\bigr]=2\lambda_3^{[XY]}-K_X,
\qquad
\E\bigl[X(12W-6)\ind(Y>y)\bigr]=R_X(y),
\end{equation*}
where continuity of $Y$ makes $Y>y$ and $Y\ge y$ equivalent
inside the expectation. Hence
\begin{equation*}
M_{3,1}^{[XY]}(x,y)
=\frac13\{xP_2(u_y)+2\lambda_3^{[XY]}+R_X(y)-K_X\},
\end{equation*}
which establishes \eqref{eq:M31main}.
All calculations have direct reverse-coordinate analogues.

\subsection{Independence covariance: all entries from finite rank sums}
Under $X\perp Y$, the $Y$-rank permutation is independent of the $X$ sample.
Conditional on that permutation, $X_{[q:n]}$ is a fixed relabeling of
$X_1,\ldots,X_n$. Hence the concomitants $X_{[1:n]},\ldots,X_{[n:n]}$ are
iid with distribution $F_X$ \citep{DavidNagaraja2003}.  Accordingly,
\begin{equation*}
\Var(\widehat\lambda_m^{[XY]})=\Var(X)\sum_{q=1}^nw_{m,q}^2,
\qquad\Cov(\widehat\lambda_2^{[XY]},\widehat\lambda_3^{[XY]})
=\Var(X)\sum_qw_{2,q}w_{3,q}.
\end{equation*}
We compute these finite sums without asymptotic approximations. Put $z_q=q-(n+1)/2$ and $h=n^2-1$. The centered power sums are
\begin{equation*}
\sum_{q=1}^n z_q=0,\quad
\sum_{q=1}^nz_q^2=\frac{nh}{12},\quad
\sum_{q=1}^nz_q^4=\frac{nh(3n^2-7)}{240}.
\end{equation*}
Since $w_{2,q}=2z_q/[n(n-1)]$ and $\sum_qz_q^2=n(n^2-1)/12$,
\begin{equation*}
\sum_qw_{2,q}^2=\frac{n+1}{3n(n-1)}=a_n.
\end{equation*}
For the third order the weight numerator is $6z_q^2-h/2$. Its squared sum is
\begin{align*}
\sum_q(6z_q^2-h/2)^2
&=36\sum_qz_q^4-6h\sum_qz_q^2+\frac{nh^2}{4}\\
&=36\frac{nh(3n^2-7)}{240}-6h\frac{nh}{12}+\frac{nh^2}{4}
=\frac{nh(n^2-4)}5.
\end{align*}
Consequently
\begin{equation*}
\sum_qw_{3,q}^2
=\frac{nh(n^2-4)}{5n^2(n-1)^2(n-2)^2}
=\frac{(n+1)(n+2)}{5n(n-1)(n-2)}=b_n.
\end{equation*}
The mixed sum vanishes exactly because $w_{2,n-q+1}=-w_{2,q}$ whereas $w_{3,n-q+1}=w_{3,q}$. This gives the two same-direction covariance blocks.

For the opposite directions, write $r_X(i),r_Y(i)$ for the sample ranks of the $i$th pair. The two complete L-comoments are
\begin{equation*}
\widehat\lambda_m^{[XY]}=\sum_{i=1}^n X_iw_{m,r_Y(i)},\qquad
\widehat\lambda_s^{[YX]}=\sum_{j=1}^n Y_jw_{s,r_X(j)}.
\end{equation*}
Their population means vanish at independence. Expanding the product and separating the two independent marginal samples yields
\begin{align*}
\E(\widehat\lambda_m^{[XY]}\widehat\lambda_s^{[YX]})
&=\sum_{i=1}^n\sum_{j=1}^n
\E(X_iw_{s,r_X(j)})\E(Y_jw_{m,r_Y(i)}).
\end{align*}
Exchangeability and the univariate L-moment rank representation give
\begin{equation*}
\E(X_iw_{s,r_X(i)})=\frac{\ell_s^X}{n},\qquad
\E(Y_iw_{m,r_Y(i)})=\frac{\ell_m^Y}{n}.
\end{equation*}
For any fixed $i$, the sum over all $j$ of $w_{s,r_X(j)}$ is zero in every realized sample; multiply by $X_i$ and take expectation:
\begin{align*}
0&=\E(X_i\sum_{j=1}^nw_{s,r_X(j)})\\
&=\frac{\ell_s^X}{n}+(n-1)\E(X_iw_{s,r_X(j)}),\quad j\ne i.
\end{align*}
Hence the off-diagonal expectations are
\begin{equation*}
\E(X_iw_{s,r_X(j)})=-\frac{\ell_s^X}{n(n-1)},\quad
\E(Y_jw_{m,r_Y(i)})=-\frac{\ell_m^Y}{n(n-1)},\quad i\ne j.
\end{equation*}
The diagonal terms contribute
$ n(\ell_s^X/n)(\ell_m^Y/n)=\ell_s^X\ell_m^Y/n$,
and the $n(n-1)$ off-diagonal terms contribute
$\ell_s^X\ell_m^Y/\{n(n-1)\}$. Their sum is
\begin{equation*}
\frac{\ell_s^X\ell_m^Y}{n}+\frac{\ell_s^X\ell_m^Y}{n(n-1)}
=\frac{\ell_s^X\ell_m^Y}{n-1}.
\end{equation*}
Together with the within-direction sums this proves Theorem~\ref{thm:indep}, including the otherwise easily overlooked nonzero cross-direction covariances.

\section{Rank-prefix computation of the exact unbiased covariance}\label{app:fast}
Here we justify the complexity assertion of Proposition~\ref{prop:fast}. The combinatorial argument uses fixed-order U-statistic incidence counting \citep{Serfling1980}; the range-sum implementation uses two-dimensional prefix tables \citep{Crow1984}. Write the $p$ kernels in one vector, with orders $m_a\in\{1,2,3\}$. For $|K|=c\le m_a$ define the uncentered incidence sums
\begin{equation}\label{eq:rawincidence}
\mathcal I_{a,K}=\sum_{\substack{I\supseteq K\\|I|=m_a}}H_{m_a}^a(Z_I),\qquad
N_a=\binom n{m_a},\quad U_a=\mathcal I_{a,\varnothing}/N_a.
\end{equation}
Then $A_{m_a,K}^a=\mathcal I_{a,K}-\binom{n-c}{m_a-c}U_a$. Consequently, only singleton, pair, and triple incidence products are required in \eqref{eq:fastidentity}. The order-one and order-two incidence sums are immediate after $O(n^2)$ pairwise evaluations.

For the order-three kernels, fix a pair $i<j$, and denote
its ordered $Y$-ranks by $\ell_Y<h_Y$. The third observation
$k$ belongs to one of the three open $Y$-rank intervals
$[1,\ell_Y-1]$, $[\ell_Y+1,h_Y-1]$, or $[h_Y+1,n]$.
Within each interval the forward kernel is an affine function
\[
H_3^{[XY]}(Z_i,Z_j,Z_k)=(aX_k+b)/3,
\qquad a\in\{1,-2,1\},
\]
where $b$ depends only on the fixed pair and on its rank
interval. One-dimensional cumulative counts and sums of $X$
in increasing $Y$-rank order therefore evaluate the forward
component of $F_{ij}$ in constant time for each pair.

The reverse component follows analogously from cumulative
counts and sums of $Y$ in $X$-rank order. The endpoints are
\emph{excluded}: neither $k=i$ nor $k=j$ has a rank in any of
the three open intervals for the relevant ranking coordinate.
Thus no separate endpoint subtraction is required.

For cross-direction products, partition the third observation
by both its $X$- and $Y$-rank intervals, giving nine open rank
rectangles. In each rectangle the two kernels have the forms
$(aX_k+b)/3$ and $(cY_k+d)/3$. Their product is
\[
\frac{acX_kY_k+adX_k+bcY_k+bd}{9},
\]
which is linear in each variable and of total degree at most
two, not quadratic in each variable. Squaring a kernel of a
single direction instead produces $X_k^2$ or $Y_k^2$.

Construct two-dimensional prefix tables on the joint rank
grid for
\begin{equation}\label{eq:prefixstats}
1,\quad X,\quad Y,\quad X^2,\quad Y^2,\quad XY.
\end{equation}
Each open rectangle contributes its count, first moments,
and quadratic moments through four prefix-table lookups.
The rectangle boundaries use the ranks of $i$ and $j$, so
both anchor observations are omitted from all nine rectangles.
Empty intervals and rectangles contribute zero. Consequently,
for each fixed pair, the vector and matrix
\begin{equation}\label{eq:prefixFQ}
F_{ij}=\sum_{k\ne i,j}\bH_3(Z_i,Z_j,Z_k),\qquad
G_{ij}=\sum_{k\ne i,j}\bH_3(Z_i,Z_j,Z_k)\bH_3(Z_i,Z_j,Z_k)^\T
\end{equation}
can be computed in constant time after prefix-table construction.
Only one-dimensional prefix sums are needed for $F_{ij}$;
the joint rank rectangles are needed for the off-diagonal,
cross-direction entries of $G_{ij}$. Its same-direction diagonal
entries can also be calculated from one-dimensional prefix
sums of squared coordinates.

The resulting singleton and total third-order incidence sums are
\begin{equation}\label{eq:incidencecounts}
\sum_{I\ni i,\ |I|=3}\bH_3(Z_I)=\frac12\sum_{j\ne i}F_{ij},\quad
\sum_{|I|=3}\bH_3(Z_I)=\frac13\sum_{i<j}F_{ij},\quad
\sum_{|I|=3}\bH_3(Z_I)\bH_3(Z_I)^\T=\frac13\sum_{i<j}G_{ij}.
\end{equation}
For the pair-level cross products in \eqref{eq:fastidentity}, use
$\mathcal I_{a,\{i,j\}}=H_2^a(Z_i,Z_j)$ for an order-two component
and $\mathcal I_{a,\{i,j\}}=F_{ij}^a$ for an order-three component.
In particular, the $c=2$ contribution to the $(3,3)$ block
is obtained from the centered pair-incidence vectors built
from $F_{ij}$ alone. The $c=3$ contribution uses the
centered triple-kernel outer products, whose uncentered sum
is the last expression in \eqref{eq:incidencecounts} and is
computed from $G_{ij}$.

The centered incidence products are formed by subtracting the appropriate multiples of $U_a$. Thus each of the three overlap levels, and every entry of the fixed-dimensional covariance matrix, is obtained with $O(n^2)$ work after constructing $O(n^2)$ prefix storage. The associated \textsf{R} implementation supplied with this article also includes marginal second L-moments so that Corollary~\ref{cor:ratios} is directly computable. Ties require an ordering convention or a separate treatment and are excluded from this complexity statement.

\section{Proof of Theorem~\ref{thm:projectionest}}\label{app:projproof}
Each summand in \eqref{eq:phihat} is an outer product, so
$\widehat\Omega_n$ is positive semidefinite. A fixed subset of
$m$ observations contributes to the sum of
$\widehat M_{m,1}(Z_i)$ exactly $m$ times; therefore
\begin{equation*}
\frac1n\sum_{i=1}^n\widehat M_{m,1}(Z_i)
=\widehat{\boldsymbol\lambda}_m,
\qquad \sum_{i=1}^n\widehat{\boldsymbol\phi}_i=0.
\end{equation*}
The rest of the proof has four steps, which separate the mean-square approximation of the estimated first projections from the law of large numbers.

\emph{Step 1 (mean-square approximation of the first projections).} Using the standard projection argument for fixed-order U-statistics \citep{Serfling1980,vanDerVaart1998}, conditional on $Z_i$, the terms in
\eqref{eq:Mhat} are a U-statistic of fixed order $m-1$ on
$n-1$ independent observations, with conditional expectation
$M_{m,1}(Z_i)$. In its conditional variance expansion, put $N=n-1$ and
$r=m-1$. For independently and uniformly selected $r$-subsets
of the $N$ remaining indices, the fraction having a nonempty
intersection is
\[
p_{N,r}=1-\frac{\binom{N-r}{r}}{\binom Nr}.
\]
For $m=2$ ($r=1$), $p_{N,1}=1/N$. For $m=3$ ($r=2$),
\[
p_{N,2}=\frac{4N-6}{N(N-1)}\le\frac4N
\qquad(N\ge2).
\]
For $m=4$ ($r=3$), the same calculation gives $p_{N,3}\le9/N$ for
$N\ge3$.
Disjoint subsets give conditionally independent kernels and
hence zero conditional covariance. For intersecting subsets,
Cauchy--Schwarz bounds the absolute covariance, after averaging
over $Z_i$, by $\E(\|\bH_m(Z_1,\ldots,Z_m)\|^2)$;
indeed the two kernel evaluations have the same unconditional
second moment. Therefore the conditional-variance expansion,
followed by expectation over $Z_i$, gives
\begin{equation}\label{eq:MhatL2}
\E\left(\left\|\widehat M_{m,1}(Z_i)-M_{m,1}(Z_i)\right\|^2\right)
\le\frac{C_m}{n-1}\E(\|\bH_m(Z_1,\ldots,Z_m)\|^2),
\end{equation}
with $C_2=1$, $C_3=4$ (for $n\ge3$), and $C_4=9$ (for $n\ge4$).

\emph{Step 2 (approximation of the projection vectors).} Write $\boldsymbol\phi_i=\boldsymbol\phi(Z_i)$ and $\Delta_n=n^{-1}\sum_{i=1}^n\|\widehat{\boldsymbol\phi}_i-\boldsymbol\phi_i\|^2$. By \eqref{eq:augIF} and \eqref{eq:phihat}, each component of $\widehat{\boldsymbol\phi}_i-\boldsymbol\phi_i$ is a multiple of $\widehat M_{m,1}(Z_i)-M_{m,1}(Z_i)$, $\widehat{\boldsymbol\lambda}_m-\boldsymbol\lambda_m$ or $\overline Z-\E(Z)$. The bound \eqref{eq:MhatL2}, the $L^2$ convergence $\widehat{\boldsymbol\lambda}_m\to\boldsymbol\lambda_m$ of fixed-order U-statistics, and $\E(\|\overline Z-\E(Z)\|^2)=O(n^{-1})$ give $\E(\Delta_n)=O(n^{-1})$, so $\Delta_n\xrightarrow{p}0$. This step uses only second moments and involves no law of large numbers.

\emph{Step 3 (law of large numbers).} The vectors $\boldsymbol\phi_1,\ldots,\boldsymbol\phi_n$ are iid with $\E(\|\boldsymbol\phi_1\|^2)<\infty$, so the ordinary law of large numbers gives $\widetilde\Omega_n=n^{-1}\sum_i\boldsymbol\phi_i\boldsymbol\phi_i^\T\to\Omega$ almost surely.

\emph{Step 4 (combination).} Let $\|A\|_F=\{\operatorname{tr}(A^\T A)\}^{1/2}$ denote the Frobenius norm. For any vectors $\boldsymbol u,\boldsymbol v$, $\|\boldsymbol u\boldsymbol u^\T-\boldsymbol v\boldsymbol v^\T\|_F\le\|\boldsymbol u-\boldsymbol v\|^2+2\|\boldsymbol u-\boldsymbol v\|\|\boldsymbol v\|$. Averaging over $i$ and applying the Cauchy--Schwarz inequality,
\begin{equation*}
\|\widehat\Omega_n-\widetilde\Omega_n\|_F\le\Delta_n+2\Delta_n^{1/2}\Bigl(n^{-1}\sum_{i=1}^n\|\boldsymbol\phi_i\|^2\Bigr)^{1/2}.
\end{equation*}
By Steps~2 and~3 the right-hand side tends to zero in probability, and hence $\widehat\Omega_n\xrightarrow{p}\Omega$.

\section{First-order covariance of L-comoment ratios}\label{subsec:ratios}
We consider the sample versions of the standardized coefficients
$\tau_m^{[XY]}$ and $\tau_m^{[YX]}$ of Section~\ref{sec:definitions},
\begin{equation}\label{eq:ratiosdef}
\widehat\tau_m^{[XY]}=\frac{\widehat\lambda_m^{[XY]}}{\widehat\lambda_2(X)},\qquad
\widehat\tau_m^{[YX]}=\frac{\widehat\lambda_m^{[YX]}}{\widehat\lambda_2(Y)},\qquad m=2,3.
\end{equation}
The denominators are the marginal sample L-scales of the measured
variables. The covariance of these nonlinear ratios is not obtained
exactly by linear transformation of the underlying moment covariance.
Following \citet{ElamirSeheult2004}, who distinguished exact covariances
of sample L-moments from Taylor-approximated standard errors of their
ratios, we use the delta method \citep{Oehlert1992,vanDerVaart1998}.
\begin{corollary}[Joint first-order ratio covariance]\label{cor:ratios}
	Assume the margins are continuous and nondegenerate and $\E(X^2+Y^2)<\infty$. Let
	\begin{equation*}
	\widehat{\boldsymbol m}_n=\bigl(\widehat\lambda_2^{[XY]},\widehat\lambda_2^{[YX]},\widehat\lambda_3^{[XY]},\widehat\lambda_3^{[YX]},\widehat\lambda_2(X),\widehat\lambda_2(Y)\bigr)^\T
	\end{equation*}
	be the vector of the four sample L-comoments and the two marginal sample L-scales, and let $\boldsymbol m=\E(\widehat{\boldsymbol m}_n)=(m_1,\ldots,m_6)^\T$. Denote their exact finite-sample covariance by $C_{M,n}$, and its leading Hoeffding term by $\Omega_M$, so that $C_{M,n}=\Omega_M/n+O(n^{-2})$. Define the map $g:\mathbb R^4\times(0,\infty)^2\to\mathbb R^4$ by
	\begin{equation*}
	g(m_1,\ldots,m_6)=\Bigl(\frac{m_1}{m_5},\frac{m_2}{m_6},\frac{m_3}{m_5},\frac{m_4}{m_6}\Bigr)^\T,
	\end{equation*}
	so that $g(\widehat{\boldsymbol m}_n)=(\widehat\tau_2^{[XY]},\widehat\tau_2^{[YX]},\widehat\tau_3^{[XY]},\widehat\tau_3^{[YX]})^\T$ by \eqref{eq:ratiosdef}. Its Jacobian at $\boldsymbol m$ is
	\begin{equation*}
	J=Dg(\boldsymbol m)=
	\begin{pmatrix}
	1/m_5&0&0&0&-m_1/m_5^2&0\\
	0&1/m_6&0&0&0&-m_2/m_6^2\\
	0&0&1/m_5&0&-m_3/m_5^2&0\\
	0&0&0&1/m_6&0&-m_4/m_6^2
	\end{pmatrix}.
	\end{equation*}
	Then
	\begin{equation}\label{eq:ratiolimit}
	\sqrt n\{g(\widehat{\boldsymbol m}_n)-g(\boldsymbol m)\}
	\xrightarrow{d}N_4(0,J\Omega_MJ^\T).
	\end{equation}
	In particular, if $\widehat J_n=Dg(\widehat{\boldsymbol m}_n)$ and $\widehat\Omega_{M,n}$ is the projection-based covariance estimator for this enlarged moment vector, then
	\begin{equation}\label{eq:ratioplugin}
	\widehat V_{\tau,n}=n^{-1}\widehat J_n\widehat\Omega_{M,n}\widehat J_n^\T,
	\qquad n\widehat V_{\tau,n}\xrightarrow{p}J\Omega_MJ^\T.
	\end{equation}
	For arbitrary ratios $R=A/B$ and $S=C/D$, write $r=\E(A)/\E(B)$, $s=\E(C)/\E(D)$, $b=\E(B)>0$, and $d=\E(D)>0$. A first-order Taylor approximation to their covariance, when defined, is
	\begin{equation}\label{eq:ratiotaylor}
	\Cov(R,S)\ \approx\ \frac{\Cov(A,C)-s\Cov(A,D)-r\Cov(B,C)+rs\Cov(B,D)}{bd},
	\end{equation}
	where the terms on the right are entries of $C_{M,n}$ and hence admit exact finite-sample covariance formulas. Equation~\eqref{eq:ratiotaylor} is \emph{not} an exact covariance identity for the ratios.
\end{corollary}
\begin{proof}
	The appended marginal L-moments are fixed-order U-statistics, so the multivariate Hoeffding central limit theorem applies \citep{Hoeffding1948,Serfling1980}. Their population L-scales are positive; applying the delta method to $g$ proves \eqref{eq:ratiolimit}. The estimator $\widehat\Omega_{M,n}$ is consistent by Corollary~\ref{cor:marginalaugmentation}, and $\widehat J_n\to_p J$; continuous mapping yields \eqref{eq:ratioplugin}. The derivative of $a/b$ is $b^{-1}(\mathrm da-r\,\mathrm db)$, and expanding the covariance of the resulting linear combinations gives \eqref{eq:ratiotaylor}.
\end{proof}
\begin{remark}[Scope of the ratio approximation]\label{rem:ratioscope}
	The ratio estimators are generally not exactly unbiased, even though their numerator and denominator moment estimators are. Likewise, inserting an exactly unbiased estimate of $C_{M,n}$ into the linearized covariance formula does not give an exactly unbiased estimate of the covariance of the nonlinear ratios. First-order linearization is sufficient for the limiting covariance in \eqref{eq:ratiolimit}. Higher-order corrections involve the third and fourth moments of the moment estimators and additional remainder control, and are not pursued here. Appendix~\ref{subsec:ratiosim} shows that the linearization can overstate the variance of $\widehat\tau_3$ noticeably under heavy-tailed margins.
\end{remark}

\subsection{Accuracy of first-order ratio approximations}\label{subsec:ratiosim}
With the notation of Section~\ref{sec:numerical} and the three designs of Section~\ref{subsec:reference}, we also examined the first-order ratio approximation of Corollary~\ref{cor:ratios}. For each ratio $\widehat\tau$ in \eqref{eq:ratiosdef}, we computed two variances. The Taylor variance is the corresponding diagonal entry of $JC_nJ^\T$, the first-order approximation of Corollary~\ref{cor:ratios} evaluated with the reference covariance $C_n$ and the Jacobian $J$ at the population moments. The Monte Carlo variance is the sample variance of $\widehat\tau$ over the $10^6$ reference replications, which serves as the true finite-sample variance. Table~\ref{tab:ratiosim} reports the ratio of the Taylor variance to the Monte Carlo variance, averaged over the two directions, which gave nearly identical values; a value above one means that the Taylor approximation overstates the variance.

\begin{table}[htbp]
	\centering
	\caption{First-order ratio approximation: ratio of the Taylor variance to the Monte Carlo variance of the sample ratios, averaged over the two directions.}
	\label{tab:ratiosim}
	\begin{tabular}{lr cc}
		\toprule
		Design & $n$ & $\tau_2$ & $\tau_3$\\
		\midrule
		Independent & 20 & 1.074 & 1.076 \\
		Independent & 30 & 1.051 & 1.050 \\
		Independent & 50 & 1.031 & 1.031 \\
		\addlinespace
		Gaussian & 20 & 0.948 & 0.992 \\
		Gaussian & 30 & 0.965 & 0.994 \\
		Gaussian & 50 & 0.978 & 0.996 \\
		\addlinespace
		GEV copula & 20 & 1.043 & 1.292 \\
		GEV copula & 30 & 1.034 & 1.226 \\
		GEV copula & 50 & 1.029 & 1.161 \\
		\bottomrule
	\end{tabular}
\end{table}

The Taylor variance is within $8\%$ of the Monte Carlo variance for $\tau_2$ in all designs and for $\tau_3$ in the independent and Gaussian designs, and the discrepancy decreases with $n$. For $\tau_3$ under heavy-tailed GEV margins, however, it overstates the variance by $29\%$, $23\%$, and $16\%$ at $n=20$, $30$, and $50$. This agrees with Remark~\ref{rem:ratioscope}: for small samples and heavy tails, higher-order corrections may be needed for ratio inference.

{\small\linespread{1.0}\selectfont

}
\end{document}